\documentclass[a4paper,11pt]{article}
\usepackage{jcappub}
\compress
\usepackage{subcaption}

\usepackage{amsmath,amssymb,amsfonts,amsthm,mathtools}
\usepackage{bm}

\usepackage{graphicx}
\usepackage{booktabs}
\usepackage{multirow}

\usepackage{microtype}
\usepackage{enumitem}
\usepackage{needspace}
\usepackage{afterpage}
\usepackage{verbatim}
\usepackage{gensymb}
\usepackage[usenames]{color}

\usepackage[citecolor=blue,colorlinks=true]{hyperref}

\makeatletter
\gdef\@fpheader{\makebox[\textwidth][r]{ET-0725A-26}\vspace{-1.0cm}}
\makeatother

\title{Impact of site-dependent noise configurations on Einstein Telescope science at low frequency}

\author[a,b,1]{Francesco Crescimbeni,\note{Corresponding author; \url{francesco.crescimbeni@uniroma1.it.}}}
\author[c,d]{Matteo Di Giovanni,}
\author[e,f]{Davide Rozza,}
\author[g]{Andrea Contu,}
\author[g]{Alessandro Cardini,}
\author[h,g]{Domenico D'Urso,}
\author[i,g]{Carlo Giunchi,}
\author[b]{Luca Naticchioni,}
\author[j,g]{Marco Olivieri,}
\author[k,l]{Mike Lindner,}
\author[l]{Andreas Rietbrock,}
\author[a,b]{and Paolo Pani}
\affiliation[a]{Dipartimento di Fisica, Sapienza Universit\`a di Roma, P.le Aldo Moro 5, 00185, Roma, Italy}
\affiliation[b]{INFN, Sezione di Roma, P.le Aldo Moro 2, 00185, Roma, Italy}
\affiliation[c]{Scuola Normale Superiore, Piazza dei Cavalieri 7, 56126, Pisa, Italy}
\affiliation[d]{INFN, Sezione di Pisa, via Filippo Buonarroti 3, 56127, Pisa, Italy}
\affiliation[e]{Universit\`a degli Studi di Milano-Bicocca, Piazza della Scienza 3, 20126 Milano, Italy}
\affiliation[f]{INFN, Sezione di Milano-Bicocca, Piazza della Scienza 3, 20126 Milano, Italy}
\affiliation[g]{INFN, Sezione di Cagliari, S.P. per Sestu km 0,700, 09042 Monserrato (CA), Italy}
\affiliation[h]{Universit\`a degli Studi di Sassari, Italy}
\affiliation[i]{Istituto Nazionale di Geofisica e Vulcanologia, Sezione di Pisa, Italy}
\affiliation[j]{Istituto Nazionale di Geofisica e Vulcanologia, Sezione di Bologna, Italy}
\affiliation[k]{Deutsches Zentrum f\"ur Astrophysik (DZA), G\"orlitz, Germany}
\affiliation[l]{Karlsruhe Institute of Technology, Geophysical Institute, Karlsruhe, Germany}

\abstract{
We investigate how representative site-dependent Newtonian noise configurations affect the scientific performance of the Einstein Telescope (ET), a proposed third-generation gravitational-wave observatory.
We compare the baseline triangular and 2L detector geometries, with arm lengths of $10\,{\rm km}$ and $15\,{\rm km}$, respectively, under three representative site-dependent noise scenarios constructed from measurements in the Euregio Meuse-Rhine region, Lausitz, and Sardinia.
The differences among the adopted sensitivities are concentrated mainly below $10\,{\rm Hz}$ and therefore have the greatest impact on science cases that rely on the early inspiral of light objects, or on massive systems whose characteristic frequencies lie in the low-frequency band.
We quantify these effects for four representative ET science cases:
(1) detection and parameter estimation of intermediate-mass black-hole binaries;
(2) post-Newtonian tests of General Relativity based on the early inspiral;
(3) black-hole ringdown spectroscopy with massive remnants;
and
(4) early-warning and sky-localization capabilities for binary neutron-star mergers.
Using Bayesian parameter estimation, we find that improved low-frequency sensitivity can significantly enhance ET's scientific performance across all four cases.
Within the configurations considered here, a 2L network in which at least one of the two L-shaped interferometers attains a noise level comparable to that assumed for the Sardinian site generally provides better performance than the triangular configurations.
Conversely, sufficiently strong degradation of the sensitivity below $10\,{\rm Hz}$ can substantially reduce, and in some regimes offset, the scientific gains expected from increasing the arm length from $10\,{\rm km}$ to $15\,{\rm km}$.
These results highlight the importance of site selection and low-frequency noise mitigation, alongside network geometry, for fully realizing ET's scientific potential in the largely unexplored frequency band below $10\,{\rm Hz}$.
}

\begin{document}

\maketitle

\section{Introduction}
\label{sec:intro}
The direct detection of gravitational waves (GWs) by the LIGO and Virgo detectors opened a new observational window on the Universe~\cite{LIGOScientific:2016aoc,TheLIGOScientific:2014jea,Virgo:2014aba,LIGOScientific:2017vwq}.
Since then, observations of compact-binary coalescences have provided new insights into the astrophysics of black holes~(BHs) and neutron stars~(NSs), the nature of gravity in the strong-field regime, and the expansion history of the Universe~\cite{LIGOScientific:2026wfs,LIGOScientific:2026ctl,LIGOScientific:2026qni,LIGOScientific:2026fcf,LIGOScientific:2026wpt}.
The sensitivity and bandwidth of current facilities, however, impose intrinsic limitations on their cosmological reach, on their ability to detect certain classes of sources, and on the accuracy with which many source properties can be reconstructed.
These limitations motivate the development of third-generation ground-based observatories, in particular the Einstein Telescope~(ET) in Europe and Cosmic Explorer~in the United States~\cite{Punturo:2010zz,Hild:2010id,ET:2019dnz,Branchesi:2023mws,ET:2025xjr,Reitze:2019iox,Evans:2021gyd}.

ET is designed to improve the strain sensitivity of current detectors over a broad frequency range and, crucially, to extend the accessible band down to frequencies of a few Hz \cite{Punturo:2010zz, Hild:2010id}.
In the following, we refer to the frequency range below approximately $10\,{\rm Hz}$ as the low-frequency band.

Access to this low-frequency band is one of the defining features of ET and a key element of its science case. It would allow compact-binary signals to remain in band for much longer than in current detectors, extend the detection horizon of massive systems, discover high-redshift binaries, and improve the reconstruction of source parameters through the observation of a larger number of inspiral cycles~\cite{Sathyaprakash:2012jk,ET:2019dnz,Branchesi:2023mws,ET:2025xjr}.
It is also particularly important for science cases relying directly on the early inspiral or on signals whose characteristic frequencies lie below approximately $10\,{\rm Hz}$.
These include the observation of intermediate-mass and high-redshift BH binaries, eccentricity measurements, precision tests of negative post-Newtonian corrections to General Relativity~(GR), the characterization of heavy BH mergers and their ringdown, the discovery and characterization of putative subsolar binaries, and the early detection and localization of binary NS systems~\cite{Sesana:2016ljz,Gair:2011ym,Meidam:2017dgf,Berti:2016lat,Akcay:2018aqj,Chan:2018csa,Nitz:2020eik,Bhagwat:2023jwv,Crescimbeni:2024cwh,Crescimbeni:2024qrq}.

Achieving the targeted low-frequency sensitivity is experimentally challenging and is one of the main reasons why ET will be built several hundred meters underground~\cite{Hild:2010id,Punturo:2010zz}.
At these frequencies, the detector noise budget receives contributions from environmental (seismic motion, Newtonian noise~\cite{Hughes:1998pe, Beccaria:1998ap, Harms:2019dqi}) and technical (suspension thermal noise, quantum radiation-pressure noise) noise sources, plus other environmental and instrumental disturbances~\cite{Hild:2010id,ET:2025DesignReport,DiGiovanni:2025oyo} associated with the detector infrastructure itself, which cannot yet be predicted reliably. The characteristics of technical noise are determined by the detector design and the technologies employed, while the amplitude and spectral structure of environmental noise depend on local geological, seismic, and environmental conditions. To this end, an underground location and the choice of an environmentally sustainable site can help not only to establish a more controlled environment, but also to reduce the magnitude of some environmental noise sources to levels at which they can be more effectively addressed and mitigated than in a surface infrastructure. In particular, Newtonian noise is one of the environmental noise sources whose impact is reduced underground but remains especially challenging.
It is therefore a major concern for ET. Newtonian noise arises from mass-density fluctuations in the vicinity of the interferometer, which induce fluctuations in the local gravitational field and hence in the forces acting on the test masses~\cite{Hughes:1998pe, Harms:2019dqi}. Unlike direct seismic ground motion, Newtonian noise cannot be mechanically isolated or shielded and must instead be mitigated by estimating and subtracting its contribution from the detector data stream a posteriori~\cite{cella}.

Consequently, the sensitivity ultimately achieved by ET may depend not only on the detector technology and geometry, but also on the characteristics of the selected site and on the effectiveness of the corresponding noise-mitigation strategies. The surrounding environment should therefore be regarded as an integral part of the detector design, rather than as a secondary engineering or geophysical consideration. Therefore, ongoing environmental measurements at the candidate sites provide essential inputs for assessing the feasibility of the target sensitivity and its potential scientific impact~\cite{DiGiovanni:2025oyo}.

Two detector geometries are currently under consideration for ET,
which emerged from broader studies of alternative detector geometries,
arm lengths, and network layouts, aimed at identifying the most promising options in terms of scientific performance, technical feasibility, and cost (see~\cite{Branchesi:2023mws,Dupletsa:2026jzs} for recent comparisons).
The first configuration consists of three nested interferometers arranged in a triangular geometry, with an opening angle of $60^\circ$ and an arm length of $10\,{\rm km}$. The second configuration consists of two geographically separated L-shaped interferometers, with an opening angle of $90^\circ$ and arms of $15\,{\rm km}$.

The two options differ not only in geometry and arm length, but also in their antenna response, geographical baseline, polarization reconstruction, and susceptibility to local noise conditions~\cite{Branchesi:2023mws,Dupletsa:2026jzs}.
A comparison of their scientific performance must therefore account jointly for detector geometry, arm length, network layout, and the site-dependent noise budget that can realistically be achieved. In particular, the nominal advantage provided by longer arms cannot be assessed independently of the low-frequency noise affecting the corresponding instruments.

Previous studies have investigated the scientific performance of alternative ET geometries and the response of different science metrics to modifications of the instrumental noise budget~\cite{Branchesi:2023mws,ET:2025xjr,Dupletsa:2026jzs}.
Other work has examined how locally measured noise at candidate sites propagates into the signal-to-noise ratio~(SNR) of compact-binary signals~\cite{DiGiovanni:2025oyo}.
These analyses establish the importance of connecting detector and infrastructure requirements to quantitative scientific metrics.
The purpose of the present work is complementary: we assess how representative site-dependent low-frequency noise configurations affect a selected set of ET science cases, while consistently accounting for the detector geometry and geographical layout.

We consider noise configurations associated with three candidate regions: the Euregio Meuse-Rhine~(EMR) region, Lausitz, and Sardinia. The adopted curves should be understood as representative sensitivity scenarios, obtained by combining the available site-specific seismic measurements with a common, simplified analytical estimate of the Newtonian noise (see Sec.~\ref{sec:noise}), rather than as site-specific Newtonian-noise models or as definitive predictions of the sensitivity that would ultimately be achieved at any location.

Indeed, the final noise budget will depend on the detailed detector design, the placement and orientation of the interferometers, further site characterization, and the mitigation techniques implemented during construction and operation.
Accordingly, our goal is neither to provide an overall ranking of the candidate sites nor to capture every scientific consideration entering the site-selection process.
Instead, we isolate the consequences of the differences among the noise curves considered here, which are concentrated mainly below approximately $10\,{\rm Hz}$, and quantify how they propagate into selected low-frequency science targets.

We study four representative science cases.
First, we investigate the detection and parameter estimation of intermediate-mass or high-redshift BH binaries, whose merger and ringdown can occur in the low-frequency band and which constitute potential targets for multiband observations~\cite{Gair:2011ym,Sesana:2016ljz,Vitale:2016rfr,Jani:2019ffg,Iacovelli:2025kwn,Berti:2026riu}.
Second, we assess parametrized tests of GR based on the post-Newtonian phasing of the early inspiral, for which the number of accumulated cycles at low frequency is a major source of constraining power~\cite{Arun:2006yw,Li:2011cg,Agathos:2013upa,Meidam:2017dgf,Roy:2025gzv}.
Third, we consider strong-field tests based on the merger and ringdown of massive BH binaries, including BH spectroscopy through the measurement of multiple quasinormal modes~(QNMs)~\cite{Dreyer:2003bv,Berti:2005ys,Gossan:2011ha,Meidam:2014jpa,Berti:2016lat,Isi:2019aib,Giesler:2019uxc,Ota:2019bzl,Bhagwat:2023jwv,Berti:2025hly}.
Finally, we study binary NS early-warning and sky-localization capabilities, which rely on detecting the inspiral sufficiently long before merger to enable rapid electromagnetic follow-up~\cite{Cannon:2011vi,Akcay:2018aqj,Chan:2018csa,Nitz:2020eik,Sachdev:2020lfd,Dupletsa:2026jzs}.

For each science case, we perform a controlled comparison among the selected triangular and 2L configurations, using Bayesian inference to estimate the measurement errors on the waveform parameters.
Our results should therefore be interpreted primarily as relative measures of the sensitivity of each scientific observable to the assumed low-frequency noise, rather than as exhaustive forecasts of ET performance.
This distinction is especially relevant because the absolute predictions may depend on source populations, waveform systematics, detector duty cycles, and the precise realization of the future observatory.
Within this scope, the analysis provides a direct mapping between site-dependent sensitivity below $10\,{\rm Hz}$ and the scientific capabilities that rely most strongly on this frequency range.

The remainder of this paper is organized as follows.
In Sec.~\ref{sec:detectors}, we describe the detector geometries, geographical configurations, and site-dependent noise curves considered in the analysis.
In Sec.~\ref{sec:methods}, we present the inference framework and the methods adopted for the different science cases.
The results for intermediate-mass BH binaries, parametrized tests of GR, BH spectroscopy, and binary NS early warning are presented in Sec.~\ref{sec:results}.
We summarize our conclusions and discuss the implications of low-frequency site-dependent noise for ET science in Sec.~\ref{sec:conclusions}.

\section{Detector networks}
\label{sec:detectors}
We model the detector response for the triangular and 2L layouts using the geographical coordinates, orientations, and site-dependent Newtonian noise curves described below. The triangular configuration consists of three nested $60^\circ$ interferometers with $10\,{\rm km}$ arms at a single site, whereas the 2L configuration consists of two geographically separated $90^\circ$ interferometers with $15\,{\rm km}$ arms. In both cases, each interferometer is assumed to employ the ET xylophone design. Refs.~\cite{ET:2019dnz,Branchesi:2023mws,ETSensitivityCurves} provide detailed descriptions of the two layouts. Here we focus on the specific site-dependent noise configurations, locations, and orientations adopted in our analysis. We consider representative configurations associated with the EMR, Lausitz, and Sardinia sites. 

For the EMR region, we use the village of Terziet in the Netherlands as a reference. 
Terziet is one of the locations investigated during the EMR site-characterization campaigns, and has been considered as a possible location for one of the vertices of the ET triangular configuration. The borehole seismometer installed there, at a depth of approximately $250\,{\rm m}$ at [$50.757^\circ$N, $5.906^\circ$E], has been active since March 2019 and covers the largest time span among all the neighboring sensors. In the representative triangular layout adopted here, the borehole lies approximately $0.4\,{\rm km}$ from vertex T2 (see Fig.~\ref{fig:europe_map}); this layout is illustrative and should not be read as a currently planned vertex position.\footnote{{The choice of TERZ as a representative location for the EMR site is also supported by recent measurements from the new EMR borehole network~\cite{EMRboreholes}. In particular, TERZ shows a remarkably stable underground seismic background: unlike stations more strongly affected by anthropogenic transients, its spectrum computed over the full data chunks differs from the transient-cleaned background spectrum by at most about $1\,{\rm dB}$.}}

For Sardinia, the seismic data are based on borehole measurements acquired at the P2 site in the Sos~Enattos region, at coordinates $[40.5233^\circ\mathrm{N},\,9.3489^\circ\mathrm{E}]$, in the vicinity of one of the proposed vertices of ET.  
The seismometer is installed at a depth of approximately $264\,{\rm m}$, and measurements have been available since October 2021. In this work, P2 is used to define a representative noise realization for Sardinia. As for the other sites, measurements from a single borehole should not be interpreted as a characterization of the entire candidate region.
Measurements are also available from a nearby site, P3, at a depth of approximately $252\,{\rm m}$. P3 is, on average, approximately 20\% quieter than P2 between $1$ and $10\,{\rm Hz}$, so the use of P2 provides the more conservative of the two available Sardinian estimates~\cite{DiGiovanni:2025oyo}.

The Lausitz region hosts five continuously operating pairs of seismic stations, each consisting of a surface station and a borehole station at a depth of $240\,{\rm m}$, jointly in operation since at least May 2026. The DZA-01 borehole, located near the town of Cunnewitz, has the longest continuous record, with measurements available since late 2022. For this work, we use data from the DZA13 borehole located at $[51.32358^\circ\mathrm{N},14.246941^\circ\mathrm{E}]$ and recorded since 2023. The installation site is about $300\,{\rm m}$ away from the nearest house, located on a meadow surrounded by agriculturally used crop fields. Local geology inferred by core probes shows Quaternary sediments with a thickness of around $59\,{\rm m}$, followed by a short transition zone of Paleogene into the granodiorite.

For each candidate site, we construct a representative triangular configuration. Our baseline 2L network instead consists of one L-shaped detector in Sardinia and one in Lausitz. These configurations are intended to isolate the interplay between the achievable
low-frequency sensitivity and the network geometry; they should not be interpreted as final engineering layouts or as a complete ranking of the candidate sites. Occasionally, we will also consider the nominal 2L sensitivity curve~\cite{ETSensitivityCurves}, to estimate the impact of the site-dependent low-frequency noise realization.
\subsection{Noise description}
\label{sec:noise}

\begin{figure*}[t]
    \centering
    \includegraphics[width=1.05\textwidth]
    {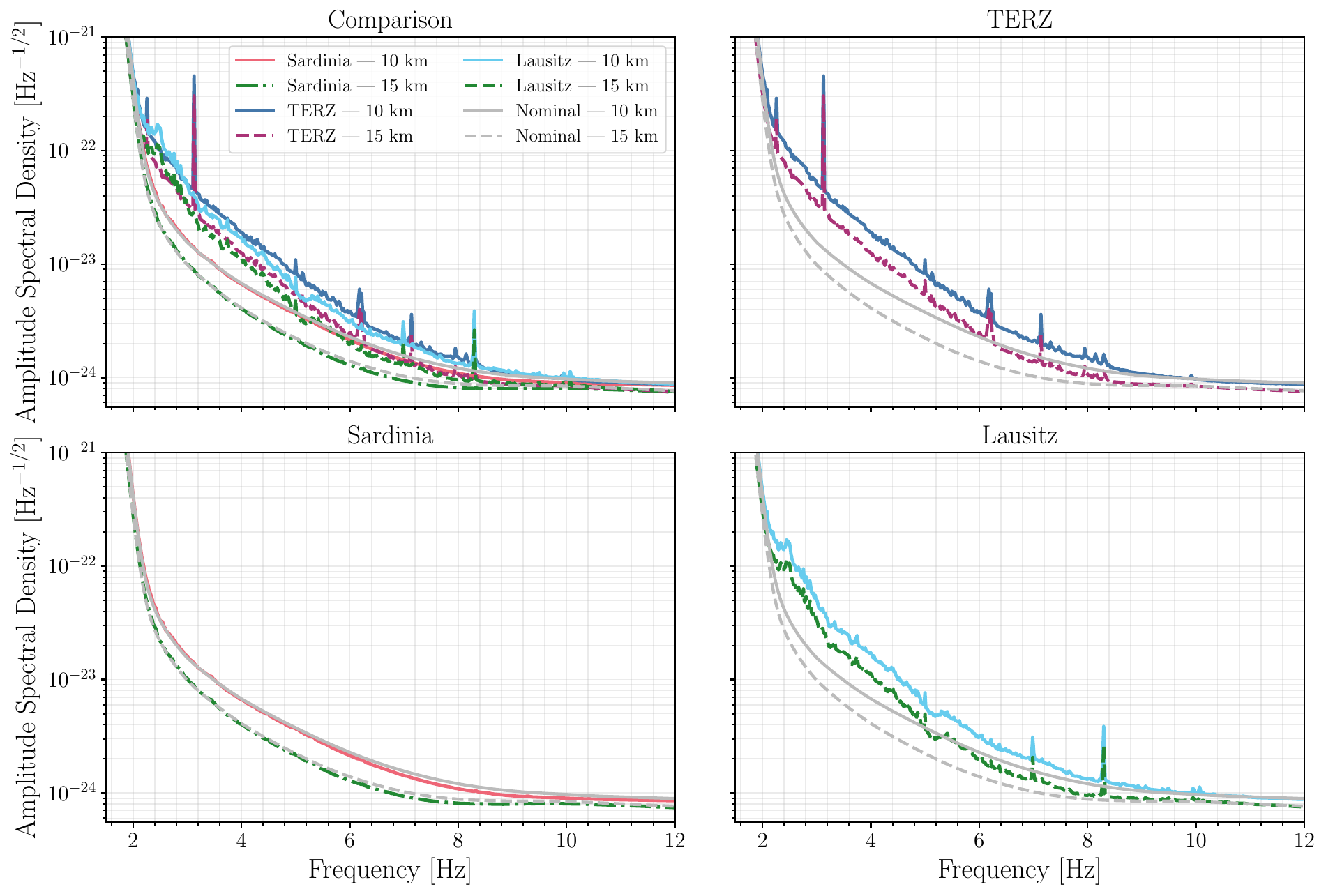}
    \caption{
        Site-dependent strain ASDs adopted in this work for the
        $10\,{\rm km}$ and $15\,{\rm km}$ configurations associated with various sites. The upper-left panel compares all the
        site-dependent curves with the corresponding nominal ET sensitivities,
        shown in gray. The remaining panels show the TERZ (top right), Sardinia (bottom left), and Lausitz (bottom right)
        configurations separately, using the same vertical scale.
        The TERZ curves correspond to the noise realization inferred from the borehole seismometer installed in the village of Terziet and should not be interpreted as a characterization of the entire EMR candidate region.
        The differences among the site-dependent, unmitigated Newtonian noise curves are concentrated mainly
        below approximately $10\,{\rm Hz}$. The Sardinia curves remain relatively
        close to the nominal targets, whereas the TERZ and Lausitz curves exhibit
        a stronger low-frequency degradation. 
        Differences in site-dependent low-frequency noise can outweigh the benefit of longer arms when comparing detectors at different sites.
    }
    \label{fig:ASDs}
\end{figure*}

The strain-noise amplitude spectral densities (ASDs) adopted in this work are shown in Fig.~\ref{fig:ASDs}. The gray curves are the
nominal ET sensitivities for the $10\,{\rm km}$ and $15\,{\rm km}$ configurations, based on
the noise models provided online in Ref.~\cite{ETSensitivityCurves} and built using GWINC~\cite{pygwinc_ascl}. 
The site-dependent curves are constructed by removing the nominal low-frequency Newtonian-noise contribution from the reference ET noise budget and replacing it with the local seismic body-wave Newtonian-noise estimate, following Ref.~\cite{DiGiovanni:2025oyo} and references therein. No separate acoustic Newtonian-noise contribution is included in the present analysis.
The Newtonian noise strain ASD is estimated as~\cite{Allocca:2021opl, Harms:2022jth, DiGiovanni:2025oyo, DiGiovanni:2026tpe}
\begin{equation}
    \widetilde{h}_{\rm NN}(f)
    =
    \frac{4\pi}{3}G\rho_0
    \frac{2\sqrt{2}}{L}
    \frac{\widetilde{x}(f)}{(2\pi f)^2},
    \label{eq:NN_ASD}
\end{equation}
where $\rho_0\approx 2.8\times 10^3\,{\rm kg}/{\rm m}^3$ is the density of the surrounding medium (assumed equal for each site), $L$ is the interferometer arm length, and $\widetilde{x}(f)$ denotes the horizontal seismic-displacement ASD used as input to the body-wave estimate.
For each site, we use the channel 1 (HH1) of the horizontal component of the corresponding borehole seismometer specified below.
This estimate assumes a spherical cavern in an infinite homogeneous medium,
an isotropic body-wave field, one third of the body-wave contribution arising
from compressional waves\footnote{Since the composition of the seismic wavefield at the sites of interest is not yet known, we assume that the displacement is equally distributed among the three seismic-wave polarizations \cite{Harms:2019dqi,Harms:2022jth}. The actual partition may vary between sites, depending on the nature and distance of seismic sources and the extent of wave scattering. Future site characterization studies will constrain this partition and inform the ET noise budget, which currently assumes the value $1/3$ adopted here.}, negligible surface-wave contributions at the detector
depth, and uncorrelated Newtonian noise between the test masses~\cite{Allocca:2021opl, Harms:2022jth, DiGiovanni:2025oyo, DiGiovanni:2026tpe}.
Within this approximation, the measured horizontal underground displacement is treated as the body-wave displacement entering Eq.~\eqref{eq:NN_ASD}.

For identical seismic input and density, the Newtonian-noise strain ASD for $15\,{\rm km}$ arms is exactly $2/3$ of that for $10\,{\rm km}$ arms.
The total strain ASD need not obey this scaling exactly, since the remaining noise contributions depend on the detector design.

Equation~\eqref{eq:NN_ASD} provides a common analytical baseline that can be applied consistently to the currently available site measurements. It should not, however, be regarded as the definitive approach to modelling Newtonian noise at ET. A dedicated effort is underway within the ET Collaboration to develop and benchmark more realistic, site-specific numerical models. To the extent permitted by the current level of site characterization, these models incorporate three-dimensional geological and geophysical information, topography, underground infrastructure, and measured properties of the seismic field. They will be progressively refined as the characterization of the candidate sites advances and are expected to yield increasingly realistic estimates of the Newtonian-noise contribution.

ET is expected to employ advanced Newtonian-noise mitigation strategies, which would introduce frequency-dependent suppression factors in Eq.~\eqref{eq:NN_ASD}~\cite{cella,Coughlin:2016tmu, Badaracco:2019vjq, Badaracco:2020qmm, vanBeveren:2023seq, Kelleter:2026pep}. 
However, the achievable level of mitigation depends strongly on the site characteristics, detector infrastructure, sensor deployment, and subtraction strategy, which have not yet been defined consistently across the candidate configurations. We therefore refrain from introducing a Newtonian-noise mitigation factor in Eq.~\eqref{eq:NN_ASD}, as any specific choice would at present be arbitrary.
The curves should therefore be regarded as representative unmitigated site-dependent sensitivity scenarios rather than predictions of the final detector performance.\footnote{
Alternative infrastructure and sensing concepts may also change the achievable
level of Newtonian-noise subtraction. In particular, configurations employing
dedicated wells and high-precision local measurements of seismic motion, tilt,
and strain are being investigated as a means of improving the reconstruction
of the seismic field near the test masses~\cite{ETPendulumSeismometers}.
At the same time, the effectiveness of such subtraction can depend on
site-specific geological structure, including the properties and geometry of
sedimentary and hard-rock layers~\cite{ETSedimentaryNN}.}

For TERZ, the seismic input is obtained from a Streckeisen STS-5A seismometer. As already noted, we use these measurements only as a proxy for a possible EMR noise realization. This comparison should therefore be interpreted with some caution, since local geology, infrastructure, and anthropogenic activity may vary substantially across the wider EMR candidate region, and measurements from a single borehole cannot characterize the site as a whole. The TERZ data also show a stronger dependence on human activity than the Sardinian measurements, with noticeably lower noise levels during nights and weekends~\cite{DiGiovanni:2025oyo}.

For P2 in Sardinia, the seismic data were acquired with a Trillium 120s SPH seismometer, with a $120\,{\rm s}$ low-frequency corner. The seismic noise measured in Sardinia is close to Peterson's New Low Noise Model~\cite{peterson1993observations} around a few Hz and, above approximately $6\,{\rm Hz}$, approaches the self-noise of the borehole sensor. The longer data set subsequently analyzed in Ref.~\cite{DiGiovanni:2026tpe} also shows that median seasonal variations have only a minor impact on the inferred ET sensitivity. In this work, we adopt the median Sardinia curve rather than seasonal extrema.

The Lausitz curves are constructed from the representative site-noise model adopted in this study, based on data acquired with a Trillium 120s PH seismometer installed at DZA13. As shown in Fig.~\ref{fig:ASDs}, the Lausitz curves are broadly similar to the TERZ curves over the frequency range relevant to our analysis, while generally exhibiting a somewhat lower noise level,
which is anyway not necessary significant given the uncertainties of the simplified Newtonian estimation model. We therefore expect the two configurations to yield qualitatively similar results for science cases primarily driven by low-frequency sensitivity, with a modest advantage for Lausitz in some regimes.

Seismic data from all three sites are analyzed consistently. The seismic displacement ASDs $\widetilde{x}(f)$ entering Eq.~\eqref{eq:NN_ASD} are obtained as the square root of the median power spectral density at each frequency, with the probabilistic power spectral density distributions computed from $120\,{\rm s}$ segments with a $50\%$ overlap. No bandpass filtering procedure is used.
The P2 dataset used in this analysis spans the period from late 2021 to late 2025. The TERZ dataset covers the same overall period, excluding the interval from mid-2023 to mid-2025 because of reported data-quality issues.\footnote{We have also verified that extending the TERZ dataset to include measurements from a new sensor operating from November~2025 to mid-2026 does not significantly modify the noise curve in the frequency range relevant to our analysis.}
Although DZA13 data are available only from 2023 onward, their time span overlaps with those of P2 and TERZ and samples the same types of seasonal variations considered for Sardinia and Terziet.

Figure~\ref{fig:ASDs} shows that the differences among the site-dependent curves are
concentrated mainly below approximately $10\,{\rm Hz}$. The Sardinia curves remain
comparatively close to the nominal sensitivities, whereas TERZ and Lausitz exhibit a more pronounced degradation at low frequency. At higher frequencies, the curves converge because the detector noise is increasingly dominated by instrumental contributions rather than by the local seismic environment.

Figure~\ref{fig:ASDs} also illustrates how differences in site noise can
outweigh the benefit of longer arms. Over a substantial part of the
few-hertz band, the $10\,{\rm km}$ Sardinia curve, which remains close
to the nominal $10\,{\rm km}$ sensitivity, lies below the
$15\,{\rm km}$ TERZ and Lausitz curves, with the crossover frequency
depending on the site and on local spectral features. This comparison
is between different seismic environments: at fixed seismic input
and density, Eq.~\eqref{eq:NN_ASD} retains its exact $1/L$ scaling.
Thus, in this regime, longer arms do not necessarily compensate for
the higher Newtonian noise at a different site, highlighting the
importance of site selection and low-frequency Newtonian-noise mitigation.

We note that the sensitivity curves adopted here do not include a recently
identified low-frequency contribution associated with the projection of vertical
suspension thermal motion onto the longitudinal interferometer direction due to the curvature of the Earth~\cite{ETVerticalThermalNoise1,ETVerticalThermalNoise2}.
For the presently considered ET low-frequency suspension concepts, this contribution may
be relevant in the $2$--$10\,{\rm Hz}$ band if not adequately mitigated.
The geometrical projection becomes more important for longer arms, and may
therefore affect quantitative comparisons between the $10\,{\rm km}$ and
$15\,{\rm km}$ configurations. Mitigation strategies for the vertical suspension thermal motion and magnetic noise are currently under
investigation. Since no consolidated noise model including these contributions is
yet available, we do not incorporate
them in the present analysis. At fixed arm length and suspension design, this geometrical coupling is common to the sites. Nevertheless, the additional noise could reduce the relative sensitivity differences among sites and alter
the corresponding science comparisons.
Our results should therefore be interpreted as a
comparison among the specific low-frequency sensitivity scenarios shown in
Fig.~\ref{fig:ASDs}, rather than as forecasts based on a complete final ET noise budget.

\subsection{Detector coordinates}
\label{sec:coordinates}

\begin{figure}[t]
    \centering
    \includegraphics[width=0.9\linewidth]{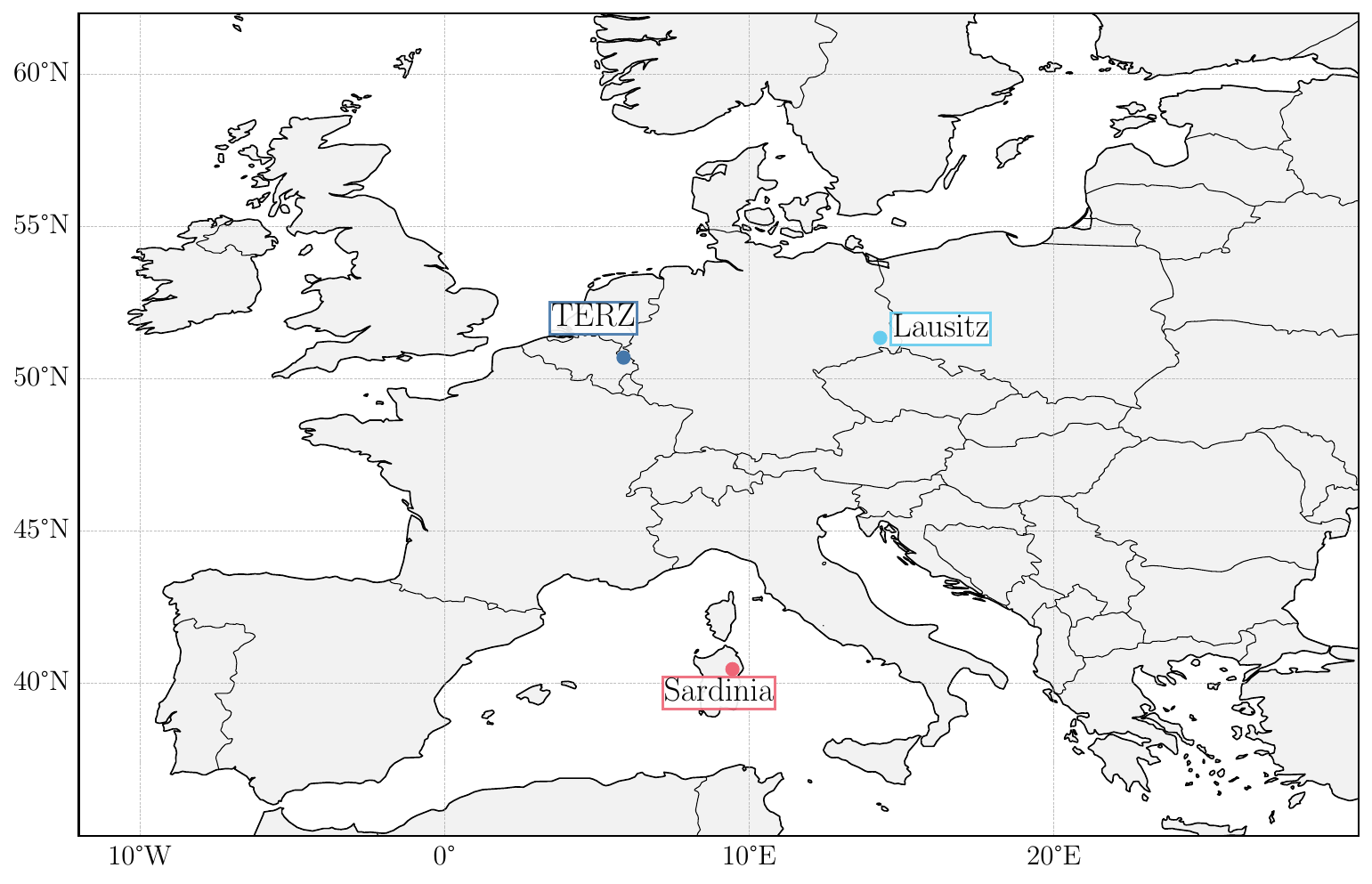}

    \includegraphics[width=0.9\linewidth]{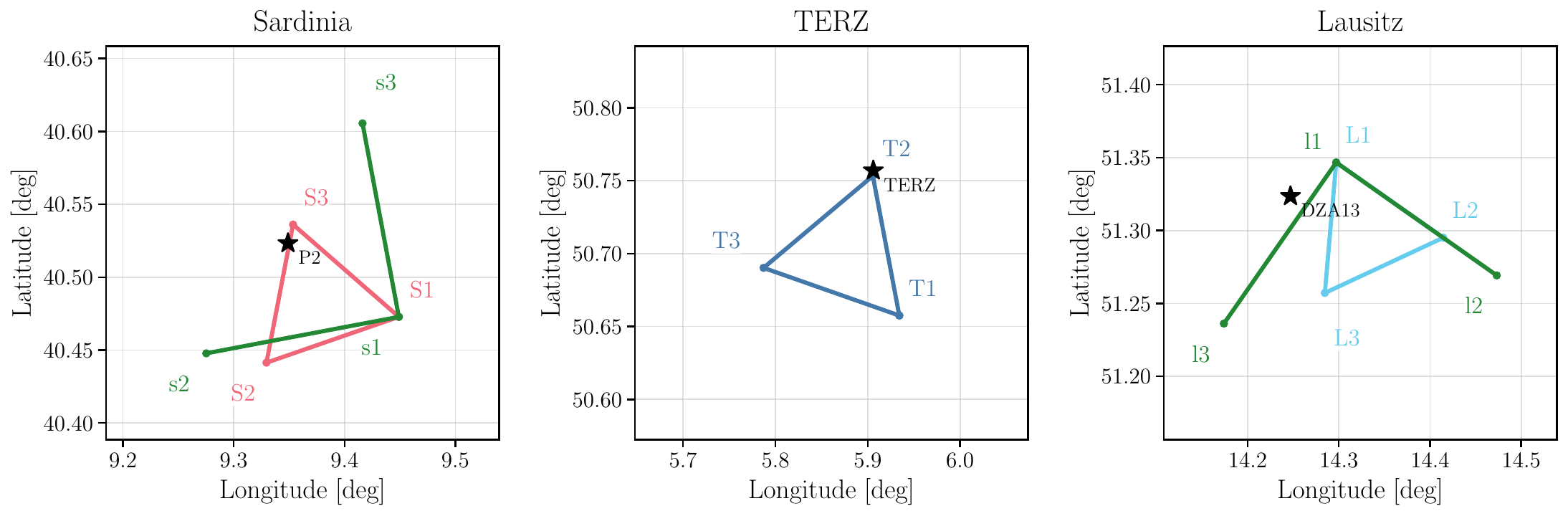}
    \caption{Top panel: locations of the different detector configurations on a map of Europe. Bottom panels: enlarged views of the three configurations, based on the coordinates listed in Tabs~\ref{tab:et_triangle_configurations} and~\ref{tab:et_l_configurations}.Black stars mark the borehole seismometers whose data enter the Newtonian-noise estimate (P2, TERZ, and DZA13), located approximately $1.5\,{\rm km}$, $0.4\,{\rm km}$, and $4.3\,{\rm km}$ from the nearest vertex of the corresponding triangular layout. The three bottom panels share the same scale, each spanning $30\,{\rm km}\times30\,{\rm km}$.}
    \label{fig:europe_map}
\end{figure}

The geographical locations and detector orientations used in the analysis are shown in
Fig.~\ref{fig:europe_map}. The corresponding vertex coordinates and arm azimuths are
reported in Tabs~\ref{tab:et_triangle_configurations}
and~\ref{tab:et_l_configurations} for the triangular and L-shaped configurations,
respectively.

For the triangular layouts, the three vertices define the detector arms and hence the
orientations of the three $60^\circ$ interferometers. For Sardinia, we use the vertex
coordinates proposed by the local site-characterization team and also adopted in
Ref.~\cite{DiGiovanni:2026tpe}. Representative triangular layouts are constructed
analogously for TERZ and Lausitz. For the 2L network, we use one $90^\circ$ interferometer in Sardinia and one in Lausitz, with the orientations given in
Table~\ref{tab:et_l_configurations}.

The top panel of Fig.~\ref{fig:europe_map} displays the three reference regions, while the lower panels show the local layouts. In all calculations, the detector tensors are built from the listed coordinates and arm azimuths, so that the antenna patterns, geographical separation, and relative orientations are consistently included in the network response.

\begin{table}[htbp]
\centering
\begin{tabular}{lrrrr}
\hline
Vertex & Latitude [deg] & Longitude [deg] & $x$-arm azimuth [deg] & $y$-arm azimuth [deg] \\
\hline
S1 & 40.472806 & 9.448954 & 198.984 & 138.986 \\
S2 & 40.441343 & 9.329487 & 79.061 & 19.062 \\
S3 & 40.536110 & 9.353495 & 319.048 & 259.045 \\
T1 & 50.657500 & 5.934200 & 100.760 & 160.652 \\
T2 & 50.753300 & 5.905500 & 219.974 & 280.782 \\
T3 & 50.690210 & 5.787140 & 340.765 & 40.066 \\
L1 & 51.346692 & 14.297119 & 325.000 & 265.000 \\
L2 & 51.295078 & 14.414558 & 204.908 & 144.908 \\
L3 & 51.257149 & 14.284634 & 85.010 & 25.010 \\
\hline
\end{tabular}
\caption{Latitude, longitude coordinates, and arm azimuths for the ET triangle ($\Delta$) configurations; see Fig.~\ref{fig:europe_map}.}
\label{tab:et_triangle_configurations}
\end{table}

\begin{table}[htbp]
\centering
\begin{tabular}{lrrrr}
\hline
Points & Latitude [deg] & Longitude [deg] & $x$-arm azimuth [deg] & $y$-arm azimuth [deg] \\
\hline
s1$\equiv$S1 & 40.472788 & 9.448977 & 190.619 & 100.665 \\
s2 & 40.447765 & 9.275188 & 10.732 & 55.755 \\
s3 & 40.605530 & 9.416177 & 280.686 & 235.663 \\
l1$\equiv$L1 & 51.346692 & 14.297119 & 325.000 & 235.000 \\
l2 & 51.269226 & 14.473178 & 144.863 & 189.863 \\
l3 & 51.236183 & 14.173929 & 55.096 & 10.096 \\
\hline
\end{tabular}
\caption{Latitude, longitude coordinates, and arm azimuths for the ET L-shaped interferometers in Sardinia and Lausitz; see Fig.~\ref{fig:europe_map}.}
\label{tab:et_l_configurations}
\end{table}

\section{Methods}
\label{sec:methods}
In this section, we describe the Bayesian inference framework adopted throughout this work.
In GW data analysis, the goal is to infer the posterior probability distribution
\(p(\vec{\theta}\mid s)\) of a set of source parameters \(\vec{\theta}\), conditioned on the observed strain data
\begin{equation}
s(t)=h(t,\vec{\theta})+n(t),
\label{output_s}
\end{equation}
where \(h(t,\vec{\theta})\) denotes the GW signal and \(n(t)\) is the detector noise, assumed here to be stationary and Gaussian.

Bayes' theorem gives the \textit{posterior} distribution of the parameters $\vec{\theta}$, i.e. the probability of observing them conditioned on the observed data $s$. Such probability can be written as:
\begin{equation}
p(\vec{\theta}\mid s)=
\frac{\mathcal{L}(s\mid\vec{\theta})\,\pi(\vec{\theta})}
{\mathcal{Z}(s)},
\label{post}
\end{equation}
where \(\mathcal{L}(s\mid\vec{\theta})\) is the likelihood, \(\pi(\vec{\theta})\) is the prior probability distribution, and \(\mathcal{Z}(s)\) is the Bayesian evidence.
For a detector network, assuming stationary Gaussian noise and statistically independent detector channels, the likelihood factorizes into the contributions from the individual instruments~\cite{Klimenko:2006rh}. Possible departures from these assumptions, including correlated and nonstationary noise, are discussed in Refs.~\cite{Janssens:2022xmo,Caporali:2025mum,Cornish:2026tjt}. The network likelihood can therefore be written as
\begin{equation}
\mathcal{L}(s\mid\vec{\theta})
\propto
\exp\left[
-\frac{1}{2}
\sum_{k\in\mathrm{detectors}}
\bigl(h_k(\vec{\theta})-s_k
\mid
h_k(\vec{\theta})-s_k\bigr)_k
\right],
\end{equation}
where $s_k$ and $h_k(\vec{\theta})$ denote, respectively, the data and the model waveform projected onto the $k$-th detector. The noise-weighted inner product for the $k$-th detector is defined as
\begin{equation}
(g\mid h)_k
=
4\,\mathrm{Re}
\int_{f_{\min}}^{f_{\max}}
\mathrm{d}f\,
\frac{\tilde{g}_k^{*}(f)\tilde{h}_k(f)}
{S_{n,k}(f)}.
\label{inner_product}
\end{equation}
Here, tildes denote Fourier transforms and $S_{n,k}(f)$ is the one-sided noise power spectral density of the $k$-th detector, namely
\begin{equation}
S_{n,k}(f)
=
\left[\mathrm{ASD}_k(f)\right]^2.
\end{equation}
Details of the detector networks and the corresponding strain ASDs are provided in Sec.~\ref{sec:detectors}. 
Throughout this work, we set $f_{\min}=2\,\mathrm{Hz}$, while $f_{\max}$ is chosen according to the specific analysis under consideration. The lower cutoff should be regarded as a nominal analysis threshold: the effective contribution of the lowest-frequency portion of the signal is determined by the shape of $S_{n,k}(f)$, which rises steeply toward low frequencies and therefore suppresses both the accumulated SNR and the corresponding parameter-estimation information.

The prior \(\pi(\vec{\theta})\) encodes the information available about the source parameters before analyzing the data.
The evidence, or marginal likelihood, is given by
\begin{equation}
\mathcal{Z}(s)
=
\int
\mathcal{L}(s\mid\vec{\theta})\,
\pi(\vec{\theta})\,
\mathrm{d}\vec{\theta},
\end{equation}
where the integral extends over the full parameter space.
Although the evidence is not the main quantity of interest in the parameter-estimation analyses considered here, it enters the normalization of the posterior and is central to Bayesian model comparison.

We perform Bayesian inference on synthetic GW data using the public software package \texttt{bilby}~\cite{BILBY} for most of the analyses, with the exception of BH spectroscopy, where we use \texttt{PyCBC Inference}~\cite{Biwer:2018osg}. The posterior distributions are sampled through the \texttt{dynesty} nested sampler~\cite{dynesty}.
Where appropriate, we reduce the computational cost of likelihood evaluations using the relative-binning technique, in which the waveform is expanded around a suitably chosen fiducial signal~\cite{Zackay:2018qdy, Narola:2023men}.

\section{Results}
\label{sec:results}
In this section we present our main results. Each subsection focuses on a specific set of science cases.

\subsection{Intermediate-mass and high-redshift binaries}
Here we assess how the low-frequency sensitivity of the different detector configurations affects the parameter estimation for intermediate-mass and high-redshift BH binaries.
Accurate measurements of the component masses and spins are particularly relevant for multiband studies, in which the early inspiral may be observed by milli-Hz or deci-Hz detectors and the late inspiral, merger, and ringdown by ET~\cite{Gair:2011ym,Sesana:2016ljz,Vitale:2016rfr,Iacovelli:2025kwn,Berti:2026riu}.
Combining observations across widely separated frequency bands can improve parameter estimation, help connect the different stages of the coalescence, and provide additional consistency tests of the source interpretation.

Measurements of the masses and spins are also important for distinguishing among possible formation channels of BH binaries: in particular, the mass spectrum, mass ratio, and spin distribution can help discriminate between isolated and dynamical formation scenarios, hierarchical mergers in dense stellar environments, and possible primordial origins~\cite{Farr:2017uvj,Mandel:2018hfr,Stevenson:2017tfq,Bartos:2016dgn,DiCarlo:2019pmf,Samsing:2017xmd,Gerosa:2018wbw,Franciolini:2021xbq,Berti:2025usa}.
These science goals rely especially strongly on the low-frequency performance of the detector, since the characteristic frequencies of the inspiral and merger shift to lower values as the total mass or the redshift increases.

\begin{figure}[t]
    \centering
    \includegraphics[width=0.48\linewidth]{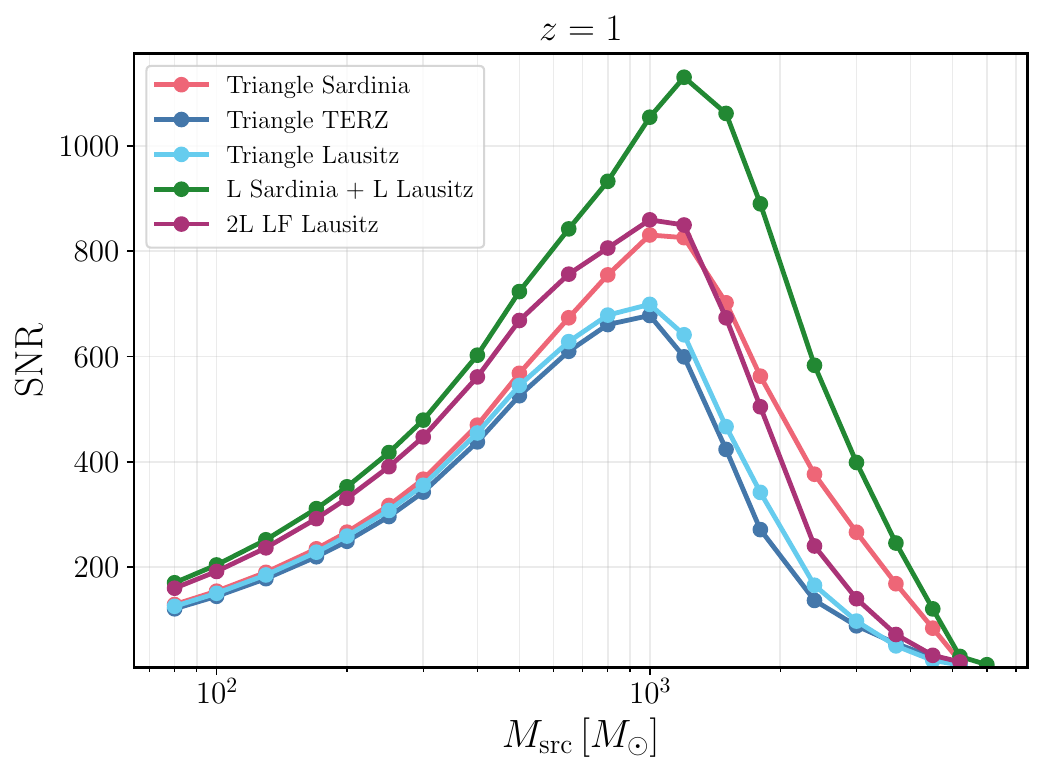}
    \includegraphics[width=0.48\linewidth]{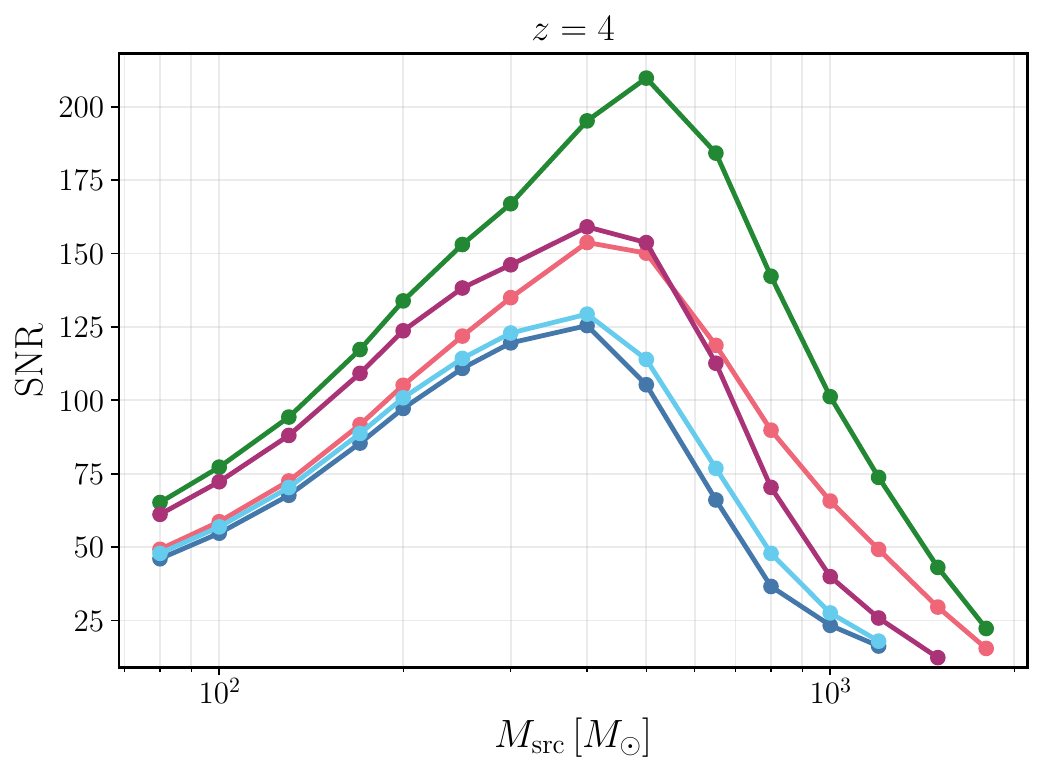}

    \includegraphics[width=0.48\linewidth]{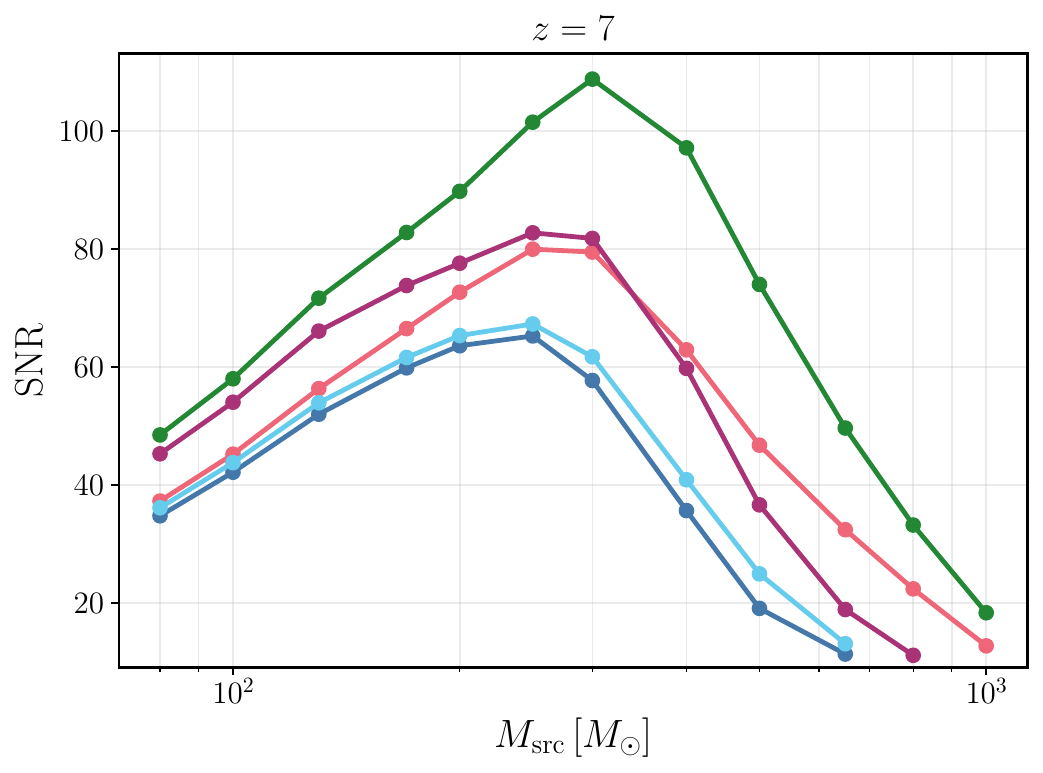}
    \includegraphics[width=0.48\linewidth]{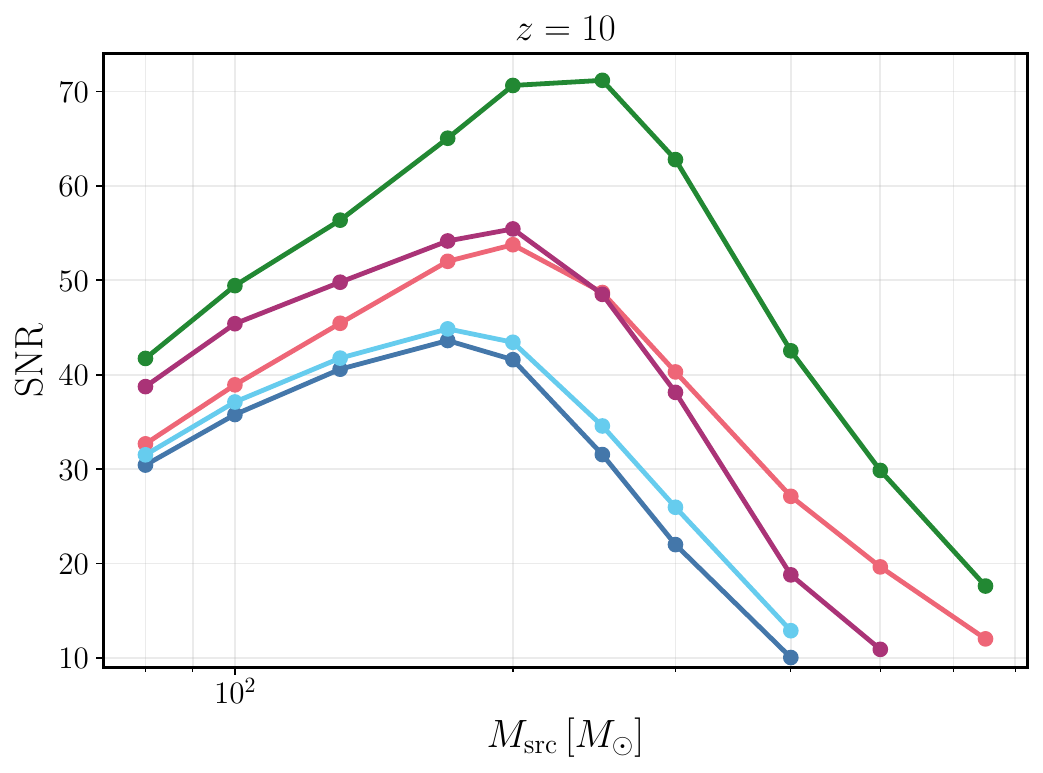}
    \caption{
    Network SNR as a function of the injected source-frame total mass for sources at $z=1$, $z=4$ (top panels), and $z=7$, $z=10$ (bottom panels), all with mass ratio $q=0.74$ and zero spins. The source sky position, inclination, and polarization are held fixed
    across all configurations to the values given in Table~\ref{tab:IMBH_injected_parameters}.} The curves correspond to the different detector sites and configurations considered in this work. The ``2L LF Lausitz'' curve represents a 2L network in which both L-shaped interferometers are assigned the low-frequency noise budget of the Lausitz configuration.
    
    \label{fig:SNRs}
\end{figure}

For this analysis, we use the frequency-domain waveform model
\texttt{IMRPhenomXHM}~\cite{Garcia-Quiros:2020qpx}, which includes higher harmonics and is therefore
appropriate for intermediate-mass binaries where the merger and ringdown can
contribute significantly to the observable signal. We restrict the analysis to quasicircular, spin-aligned binaries and neglect precession and eccentricity. For the high-mass systems considered here, the observable signal is dominated by the late inspiral, merger, and ringdown, whereas the early inspiral contributes comparatively less to the total SNR.
In Fig.~\ref{fig:SNRs}, we show the network SNR as a function of the source-frame total mass for representative redshift values. Two main trends emerge.
\begin{enumerate}
    \item
    At relatively low masses, all triangular configurations yield comparable SNRs, since the signal probes frequencies at which the corresponding sensitivity curves are similar.
    At larger masses, instead, an increasing fraction of the signal shifts into the low-frequency band, and the improved low-frequency sensitivity associated with the Sardinia site leads to
    an SNR enhancement reaching approximately a factor of $2$ in the most favorable part of the mass range.\footnote{As a reference, within a
Euclidean approximation, an SNR improvement by a factor of $2$ would correspond to an
increase in the accessible volume by a factor of $8$; this scaling is only illustrative for the cosmological sources considered here.}
    The source-frame mass at which these differences become relevant decreases with the redshift, because cosmological redshift moves the observed signal toward lower frequencies.
    \item
    The 2L configuration, consisting of one L-shaped interferometer in Sardinia and one in Lausitz,
    always provides the largest SNR, consistently with the advantage expected from its longer arms~\cite{Branchesi:2023mws}. However, Fig.~\ref{fig:SNRs} reveals a hierarchy that depends on the complete detector configuration and site-dependent noise rather than on geometry and arm length alone. Whenever at least one of the two L-shaped interferometers achieves a low-frequency sensitivity comparable to that assumed for Sardinia, the 2L network outperforms all the triangular configurations considered here over essentially the entire mass range.
    
    This conclusion, however, is conditional on the achievable low-frequency noise. The ``2L LF Lausitz'' configuration, in which both $15$-${\rm km}$ interferometers are assigned the low-frequency noise budget of the Lausitz configuration, shows that the advantage associated with the longer arms can be substantially reduced when the sensitivity below approximately $10\,{\rm Hz}$ diminishes. Once the signal becomes dominated by the lowest-frequency portion of the ET band, the reduced-sensitivity 2L network can be overtaken by the Sardinia triangle, while generally remaining more sensitive than the TERZ and Lausitz triangles. This demonstrates that the comparison between $10$-${\rm km}$ triangular and $15$-${\rm km}$ 2L layouts cannot be made independently of the low-frequency sensitivity actually achievable at the selected sites.
\end{enumerate}

\begin{figure}[t]
    \centering
    \includegraphics[width=0.7\linewidth]{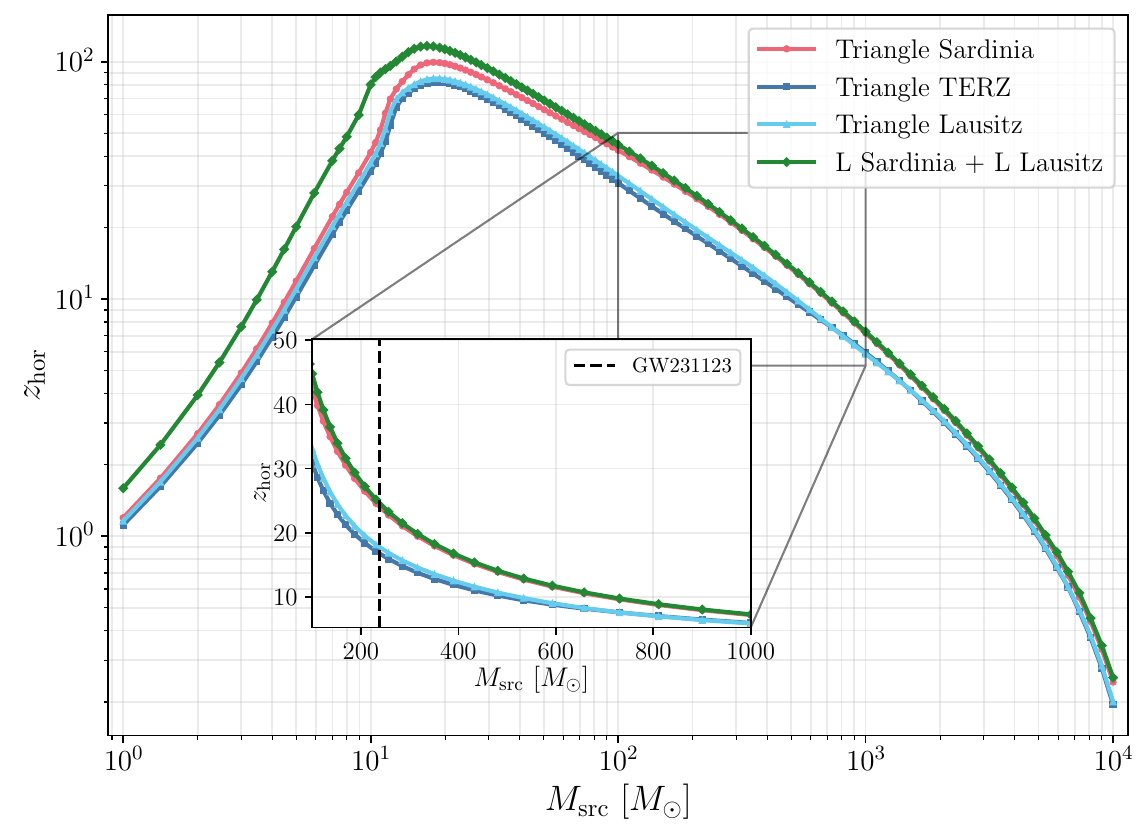}
    \caption{
    Redshift horizon for equal-mass, nonspinning binaries as a function of the source-frame total mass, assuming a network SNR threshold of $8$ for all the considered networks.
    }
    \label{fig:horizon}
\end{figure}

In Fig.~\ref{fig:horizon}, we show the redshift horizon as a function of the source-frame total mass, following a procedure similar to that of Ref.~\cite{Dupletsa:2026jzs}, with the main difference being that, in our 2L configuration, the two interferometers are assigned different power spectral densities.
For this calculation, we consider equal-mass, nonspinning binaries, and we generate the waveform with \texttt{IMRPhenomD}~\cite{Husa:2015iqa, Khan:2015jqa}, which contains only the $(\ell,|m|)=(2,2)$ mode. For each source-frame total mass, we maximize the network SNR over the source sky position, inclination, and polarization by using the \texttt{basinhopping} algorithm implemented in the \texttt{scipy} library~\cite{2020SciPy-NMeth}. The horizon redshift is then defined as the largest redshift at which an optimally located and oriented source reaches the adopted network-SNR threshold, which we set to $\rho_{\rm th}=8$, consistently with Ref.~\cite{Dupletsa:2026jzs}. At each trial redshift, the source-frame component masses are converted to detector-frame masses through the standard rescaling: 
\begin{equation}
m_i^{\rm det}=(1+z)m_i^{\rm src}.
\label{det_m_rescaling_z}
\end{equation}
The horizon is finally obtained by numerically solving:

\begin{equation}
\rho_{\rm net}\left(M_{\rm src},z_{\rm hor}\right)=\rho_{\rm th}
\end{equation}

Consistently with the trends discussed above, the different triangular configurations have nearly identical horizons at low masses, where the signal is dominated by frequencies for which the corresponding sensitivity curves are comparable.
At larger masses, the improved low-frequency sensitivity associated with the Sardinia configuration increases the redshift horizon by up to approximately a factor of $2$. The inset of Fig.~\ref{fig:horizon} indicates, for reference, the remnant mass inferred for GW231123, which lies in the intermediate-mass BH regime~\cite{LIGOScientific:2025rsn}.
Interestingly, from comparable masses onward, the Sardinia triangular configuration
approaches the redshift horizon of the 2L network.
This result highlights that, in this mass range, the achievable low-frequency noise level can be at least as important as the increase in arm length from $10\,{\rm km}$ to $15\,{\rm km}$.

For sources at $z=1$, we perform a Bayesian inference analysis using \texttt{bilby}~\cite{BILBY}. The injections are generated in zero noise with \texttt{IMRPhenomXPHM}, and the corresponding parameters are listed in Table~\ref{tab:IMBH_injected_parameters}, while the prior ranges are reported in Table~\ref{tab:IMBH_priors} in the Appendix. For the injected values, we vary the source-frame total mass over the range considered, while keeping the mass ratio and angular parameters fixed. In particular, we set $q \equiv m_2/m_1=0.74$, so that the source-frame component masses are $m_1^{\rm src}=M_{\rm tot}^{\rm src}/(1+q)$ and $m_2^{\rm src}=qM_{\rm tot}^{\rm src}/(1+q)$. The corresponding detector-frame masses are obtained through Eq.~\eqref{det_m_rescaling_z}, assuming a $\Lambda$CDM cosmology~\cite{Efstathiou:1990xe}. We adopt an inclination angle $\iota=0.2~{\rm rad}\simeq 11.46^\circ$. Although this nearly face-on configuration can enhance the degeneracy between luminosity distance and inclination~\cite{Usman:2018imj}, this degeneracy can be partially broken by a network of third-generation detectors, including when the contribution of higher-order modes is modest~\cite{Mascioli:2025cnw}.
The duration of each data segment is set by the time to merger computed from the injected detector-frame masses and the adopted low-frequency cutoff at $2\,{\rm Hz}$.

\begin{table}[t]
\centering

\begin{subtable}{0.48\linewidth}
\centering
\begin{tabular}{cc}
\toprule
Intrinsic parameter & Injected value \\
\midrule
\(M_{\rm tot}^{\rm src}\) & varied \\
\(q=m_2/m_1\) & \(0.74\) \\
\(\chi_1\) & \(0.7\) \\
\(\chi_2\) & \(0.4\) \\
\(\theta_1\) & \(0\) \\
\(\theta_2\) & \(0\) \\
\(\phi_{12}\) & \(0\) \\
\(\phi_{JL}\) & \(0\) \\
\bottomrule
\end{tabular}
\caption{Intrinsic parameters.}
\label{tab:IMBH_injected_intrinsic}
\end{subtable}
\hfill
\begin{subtable}{0.48\linewidth}
\centering
\begin{tabular}{cc}
\toprule
Extrinsic parameter & Injected value \\
\midrule
\(z\) & \(1.0\) \\
\(d_L\) & \(d_L(z)\) \\
\(\iota\) & \(11.46^\circ\) \\
\(\psi\) & \(0\) \\
\(\phi_c\) & \(1.3\) \\
\(t_c\) & \(1126259462.43~{\rm s}\) \\
\(\alpha\) & \(1.097\) \\
\(\delta\) & \(-0.13\) \\
\bottomrule
\end{tabular}
\caption{Extrinsic parameters.}
\label{tab:IMBH_injected_extrinsic}
\end{subtable}

\caption{
Injected parameters used in the Bayesian inference analysis of intermediate-mass BH binaries with \texttt{bilby}. The source-frame total mass $M_{\rm tot}^{\rm src}$ is varied, while the mass ratio, spin vectors, orientations, sky location, inclination, polarization angle, phase, and coalescence time are held fixed. The detector-frame component masses are obtained by redshifting the corresponding source-frame masses, while the luminosity distance is determined from the injected redshift using the $\Lambda$CDM model.}
\label{tab:IMBH_injected_parameters}
\end{table}

We sample over the following 11 parameters:
\begin{equation}
\vec{\theta}
=
\left\{
\mathcal{M}^{\rm det},\eta,\chi_1,\chi_2,d_L,\iota,t_c,\alpha,\delta,\psi,\phi_c
\right\},
\end{equation}
where $\mathcal{M}^{\rm det}$ is the detector-frame chirp mass,
$\eta=q/(1+q)^2$ is the symmetric mass ratio, $\chi_1$ and $\chi_2$
are the dimensionless aligned-spin parameters, $d_L$ is the luminosity
distance, $t_c$ and $\phi_c$ are the coalescence time and phase,
respectively, $\alpha$ and $\delta$ are the right ascension and
declination, and $\psi$ is the polarization angle.

The main results for the measurement uncertainties on the detector-frame chirp mass and effective spin are presented in Fig.~\ref{fig:chirpd_chieff}, while additional results are reported in the Appendix. 
We find that the triangular configurations at the Lausitz and TERZ sites yield broadly similar results, with Lausitz generally performing slightly better, while the Sardinia triangle performs overall better, especially as the source mass increases, and thus the signal is shifted to lower frequencies. 
For high-mass systems, $M_{\rm src}\sim {\rm few}\times10^3\,M_\odot$, the Sardinia triangle reduces the measurement errors relative to the TERZ and Lausitz triangles by factors of approximately $1.5$--$3$ for the detector-frame chirp mass and $1.5$--$2$ for the effective spin.
The 2L configuration generally provides the best performance, with errors smaller by approximately $20$-$50\%$ than those of the Sardinia triangle and by up to a factor of a few relative to the TERZ and Lausitz triangles. 
The chirp mass is also measured more accurately at smaller masses, because it benefits from the longer low-frequency inspiral.

\begin{figure}[t]
    \centering
    \includegraphics[width=0.48\linewidth]{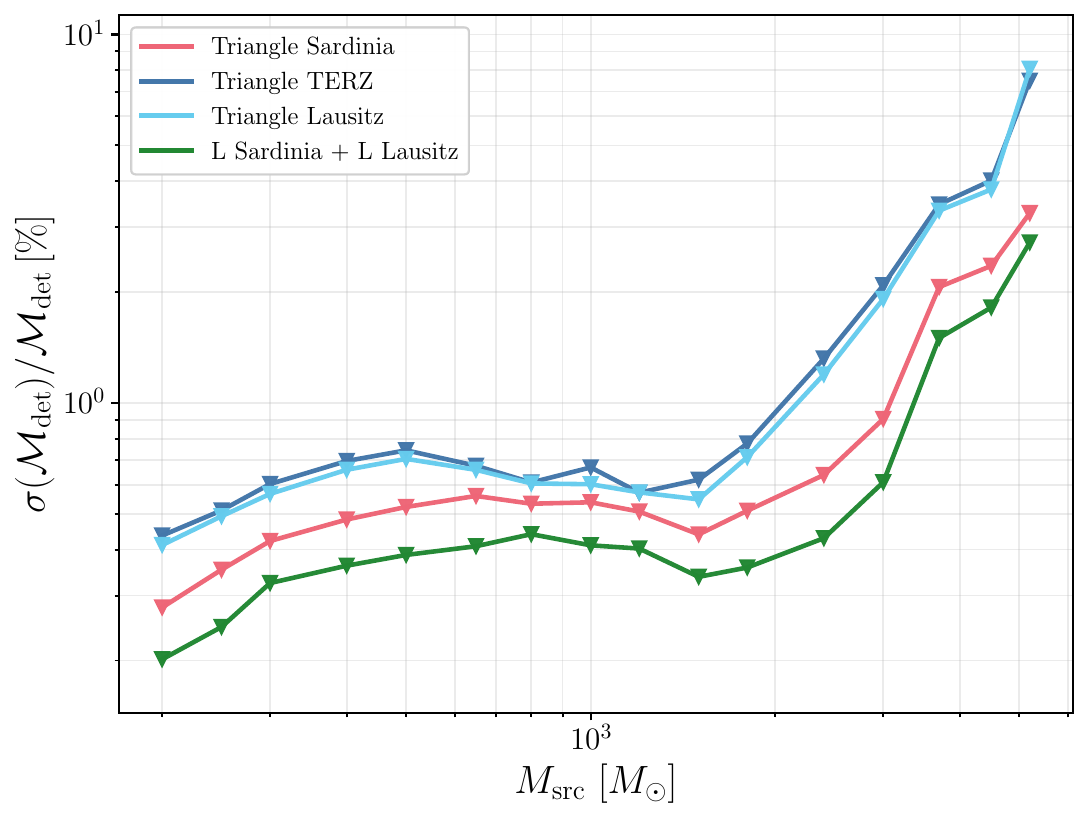}
    \includegraphics[width=0.48\linewidth]{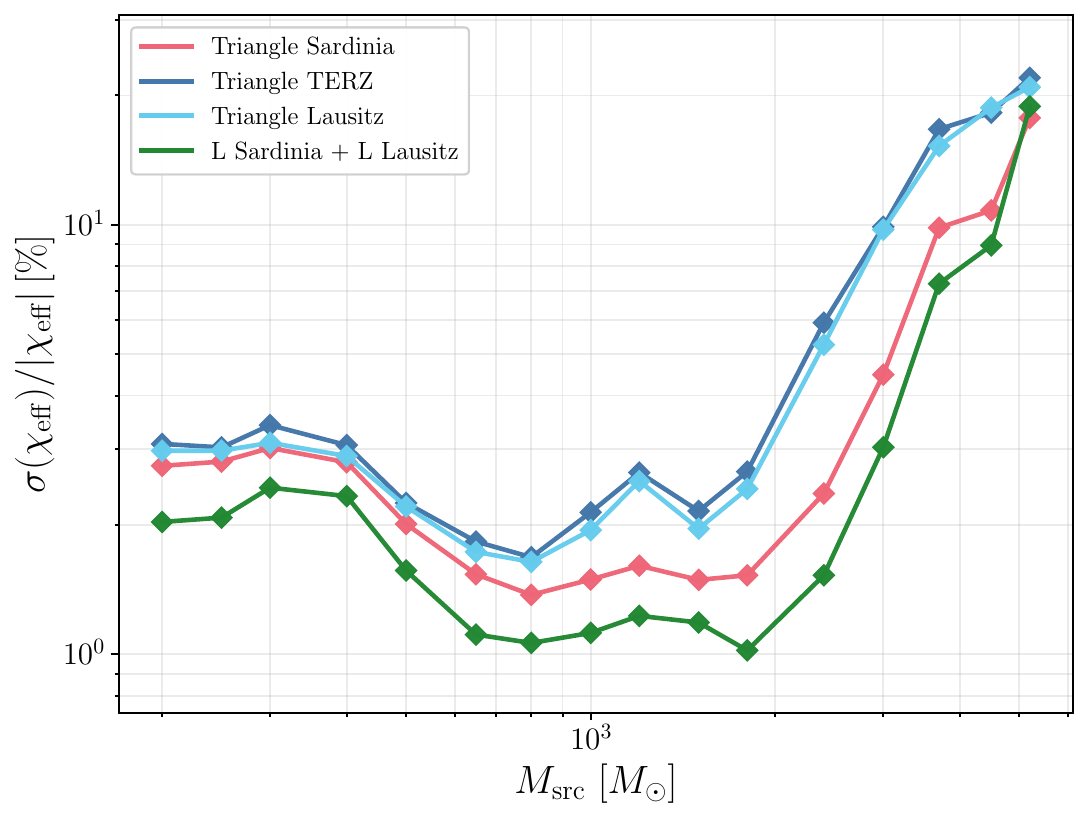}
    
    \caption{Detector-frame chirp-mass relative error as a function of the source-frame total mass (left panel) and effective-spin (right panel) relative error for \(z=1\), for the four different detector configurations.}
    \label{fig:chirpd_chieff}
\end{figure}

\subsection{Parametrized tests of gravity}
\label{sec:parametrized_tests}

We now investigate the impact of the detector network on parametrized tests of GR. We follow the logic of the TIGER framework~\cite{Agathos:2013upa, Meidam:2017dgf, Roy:2025gzv}, in which possible deviations from GR are introduced directly in the frequency-domain phase of the waveform. The GR hypothesis corresponds to all deviation parameters being equal to zero. The analysis is therefore a null test: signals are injected assuming GR, while the recovery allows one beyond-GR coefficient at a time to vary.

The injected signals are generated with the same intrinsic and extrinsic
parameters of Tab.~\ref{tab:IMBH_injected_parameters}. In this section we focus on
sources at \(z=4\), which allows us to explore the mass dependence of the
constraints while keeping the signals sufficiently loud across the considered
detector configurations.

For each injected signal and for each detector configuration, we perform
independent Bayesian inference runs in which a single beyond-GR parameter is
allowed to vary, while all the other beyond-GR coefficients are fixed to their
GR value. This is the standard single-parameter strategy used in current LIGO-Virgo-KAGRA analyses: the deviation parameters are switched on one at a time to
avoid the loss of constraining power caused by strong correlations among
multiple deviation coefficients.

In the notation used below, the sampled deviation parameter is denoted
generically by \(\delta p_i\)~\cite{Roy:2025gzv}, with 
\begin{equation}
\delta p_i
\in
\left\{
\delta \phi_i,\,
\delta \beta_i,\,
\delta \alpha_i
\right\},    
\end{equation}
where \(\delta \phi_i\), \(\delta \beta_i\), and
\(\delta \alpha_i\) denote, respectively, the inspiral, intermediate, and
merger-ringdown deviations. The TIGER framework applies the inspiral deviations
to the PN phase coefficients, while the post-inspiral parameters modify the
phenomenological coefficients calibrated to numerical relativity in the
intermediate and merger-ringdown regimes~\cite{Roy:2025gzv}.

The sampled GR parameters are
\begin{equation}
\vec{\theta}_{\rm GR}
=
\left\{
\mathcal{M}^{\rm det},
\eta,
\chi_1,
\chi_2,
d_L,
\iota,
t_c,
\alpha,
\delta,
\psi,
\phi_c
\right\}.    
\end{equation}
When testing a given beyond-GR coefficient, the full sampled parameter vector is
therefore
\begin{equation}
\vec{\theta}
=
\vec{\theta}_{\rm GR}
\cup
\left\{
\delta p_i
\right\}.    
\end{equation}
For each beyond-GR parameter, we summarize the constraint using half the width of the central $90\%$ credible interval,
\begin{equation}
\sigma_{90\%}(\delta p_i)
=
\frac{1}{2}
\left[
\left(\delta p_{i,50\%}-\delta p_{i,5\%}\right)
+
\left(\delta p_{i,95\%}-\delta p_{i,50\%}\right)
\right],  
\end{equation}
where $\delta p_{i,5\%},\delta p_{i,50\%},\delta p_{i,95\%}$ are respectively the 5th, 50th, and 95th credible percentiles of the marginalized posterior of the deviation parameter $\delta p_{i}$.
Since the injected value is always zero, smaller values of \(\sigma_{90\%}\)
correspond to tighter constraints on possible departures from GR.

\subsubsection{Inspiral}
\label{sec:parametrized_inspiral}

\begin{figure}
    \centering
        \includegraphics[width=1.05\linewidth]{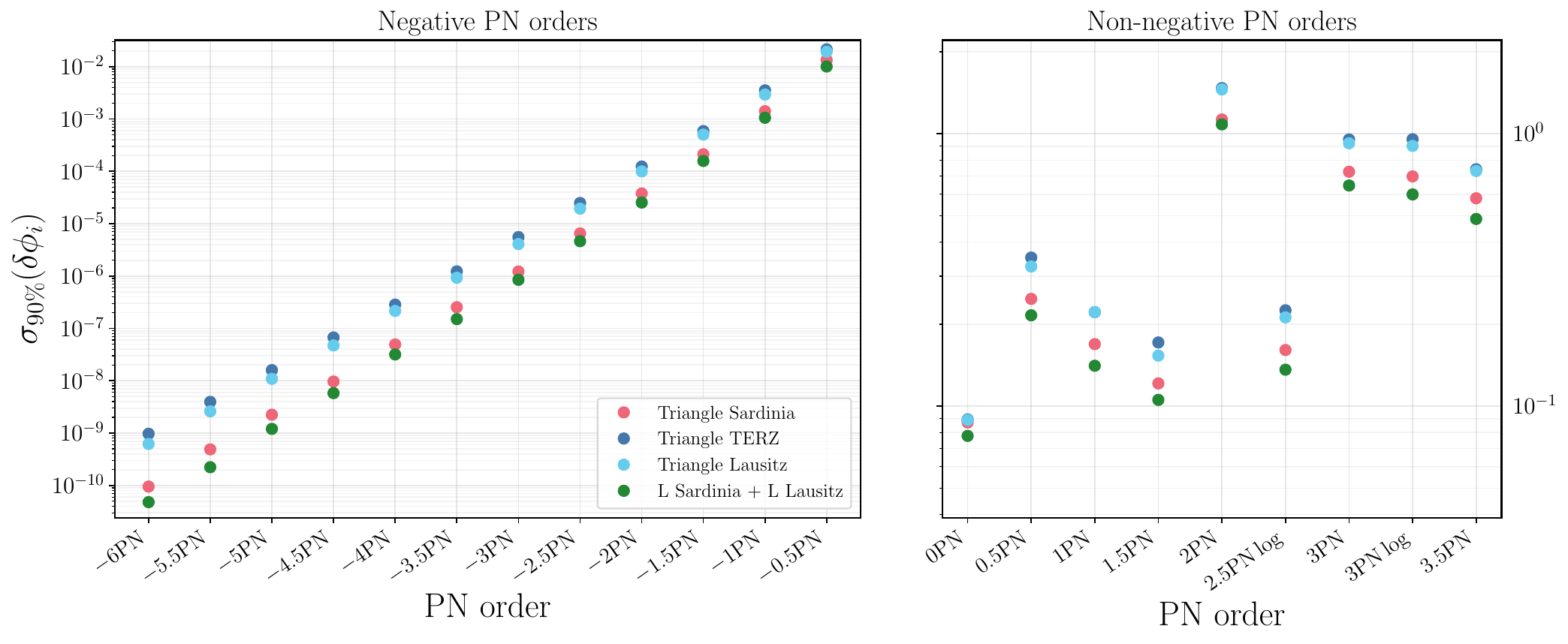}
    \caption{Half-width of the central $90\%$ credible interval, $\sigma_{90\%}$, of the parametrized PN-deviation posteriors for sources at $z=4$, for the detector configurations considered in this work. The left and right panels correspond to negative and non-negative PN terms, respectively (note the widely different vertical scales in the two panels).}
    \label{fig:PN_tests}
\end{figure}

In the inspiral regime, the frequency-domain phase can be written schematically
as
\begin{equation}
\Psi_{\rm insp}(f)=2\pi ft_c-\phi_c-\frac{\pi}{4}+\frac{3}{128\eta}v^{-5}\left[\sum_i\phi_iv^i+\sum_i\phi_i^{\ell}v^i\log v\right].
\end{equation}
where $v = (\pi M^{\rm det} f)^{1/3}$ is the binary velocity and $M^{\rm det}$ is the detector-frame total mass. Parametrized deviations are introduced by deforming the PN coefficients,
schematically as~\cite{Agathos:2013upa}
\begin{equation}
\phi_i \rightarrow \phi_i^{\rm GR}
\left(1+\delta {\phi}_i\right),
\qquad
\phi_i^{\ell} \rightarrow
\phi_i^{\ell, \rm GR}
\left(1+\delta {\phi}_i^{\ell}\right),    
\end{equation}
for coefficients that are non-zero in GR. For PN orders whose GR coefficient
vanishes, the corresponding deviation is treated as an absolute correction
rather than as a fractional one.
Although negative-PN terms are absent from the vacuum-GR phasing of quasicircular compact binaries, including them is well motivated in a theory-agnostic test. Several physical mechanisms can modify the dynamics predominantly at large orbital separations and therefore enter at an effectively negative PN order. Examples include additional channels of radiation, such as dipolar emission, modifications of the binary dynamics associated with extra gravitational degrees of freedom or slowly varying couplings, and environmental or kinematic effects (see~\cite{Barausse:2014tra} for an overview). Negative-PN coefficients thus provide a generic parametrization of departures from the standard inspiral that are enhanced in the weak-field, low-frequency regime. Extending the analysis to these terms is especially relevant for ET, whose sensitivity down to a few Hz provides access to a frequency range that is largely unavailable to current ground-based detectors.

In the inspiral sector, we vary one PN deviation parameter at a time, while all
the other beyond-GR coefficients are fixed to zero. The set of inspiral
parameters considered in the recovery is

\begin{equation}
\left\{\delta{\phi}_{i}\right\}_{i=-12}^{4}
\cup
\left\{
\delta{\phi}_{5\ell},
\delta{\phi}_{6},
\delta{\phi}_{6\ell},
\delta{\phi}_{7}
\right\},
\end{equation}
whereas, as in the standard LIGO-Virgo-KAGRA analyses~\cite{LIGOScientific:2026fcf, LIGOScientific:2026oim}, we do not present results for the 2.5PN term, which is completely degenerate with the coalescence phase. The inspiral coefficients are expected to be mostly constrained by the number
of inspiral cycles accumulated in band. Therefore, for fixed redshift and detector configuration, the constraints are generally expected to degrade as the source-frame total mass increases and the inspiral portion of the signal
moves to lower frequencies and becomes shorter in band.

Figure~\ref{fig:PN_tests} shows the constraints on parametrized deviations in the inspiral phase for a representative source with $M_{\rm src}=40\,M_\odot$ at $z=4$, while all other parameters are fixed to the values listed in Table~\ref{tab:IMBH_injected_parameters}. The most prominent feature is the strong dependence of the constraints on PN order. Negative-PN corrections are constrained much more tightly than non-negative-PN terms, with the bounds becoming progressively stronger toward increasingly negative orders. This behavior follows directly from their frequency dependence. A correction entering at PN order $n$ contributes schematically to the Fourier-domain phase as
\begin{equation}
    \delta\Psi_n(f)\propto v^{-5+2n}.
\end{equation}
Negative-PN terms therefore grow rapidly toward low frequencies and accumulate a large dephasing over the long early inspiral. Their measurement is consequently particularly sensitive to the detector performance below approximately $10\,{\rm Hz}$.

The comparison among detector configurations confirms this expectation. 
The triangular configurations at the TERZ and Lausitz sites provide broadly similar constraints, with Lausitz typically yielding slightly tighter bounds, whereas the Sardinia triangle yields systematically tighter bounds throughout the negative-PN sector. The 2L configuration gives the strongest constraints overall. The relative separation among the configurations is largest for the most negative PN orders, for which most of the constraining power comes from the lowest-frequency portion of the signal.
The difference can be as large as an order of magnitude for $-6{\rm PN}$, which corresponds to certain environmental effects~\cite{Cardoso:2019rou}.
As the PN order increases toward zero, the deformation becomes less strongly weighted toward low frequencies and the differences among the configurations decrease.

For the non-negative-PN coefficients, the hierarchy among the detector configurations remains broadly the same, but the relative improvements are generally smaller (notice the different vertical scales in the panels of Fig.~\ref{fig:PN_tests}) and the dependence on PN order is not monotonic. These terms receive an increasingly important contribution from the later inspiral, where the assumed sensitivity curves are more similar. Moreover, their constraints are affected by correlations with the intrinsic binary parameters and by partial degeneracies with the coalescence time and phase. The numerical widths at different PN orders should also not be compared as direct measures of physical relevance, since the deformation parameters have different normalizations and are treated as absolute corrections whenever the corresponding GR coefficient vanishes. The robust conclusion is instead the relative comparison among detector configurations at fixed PN order: preserving the best possible low-frequency sensitivity produces the largest scientific gain for negative-PN tests.

\subsubsection{Intermediate}
\label{sec:parametrized_intermediate}

\begin{figure}
    \centering
    \includegraphics[width=0.9\linewidth]{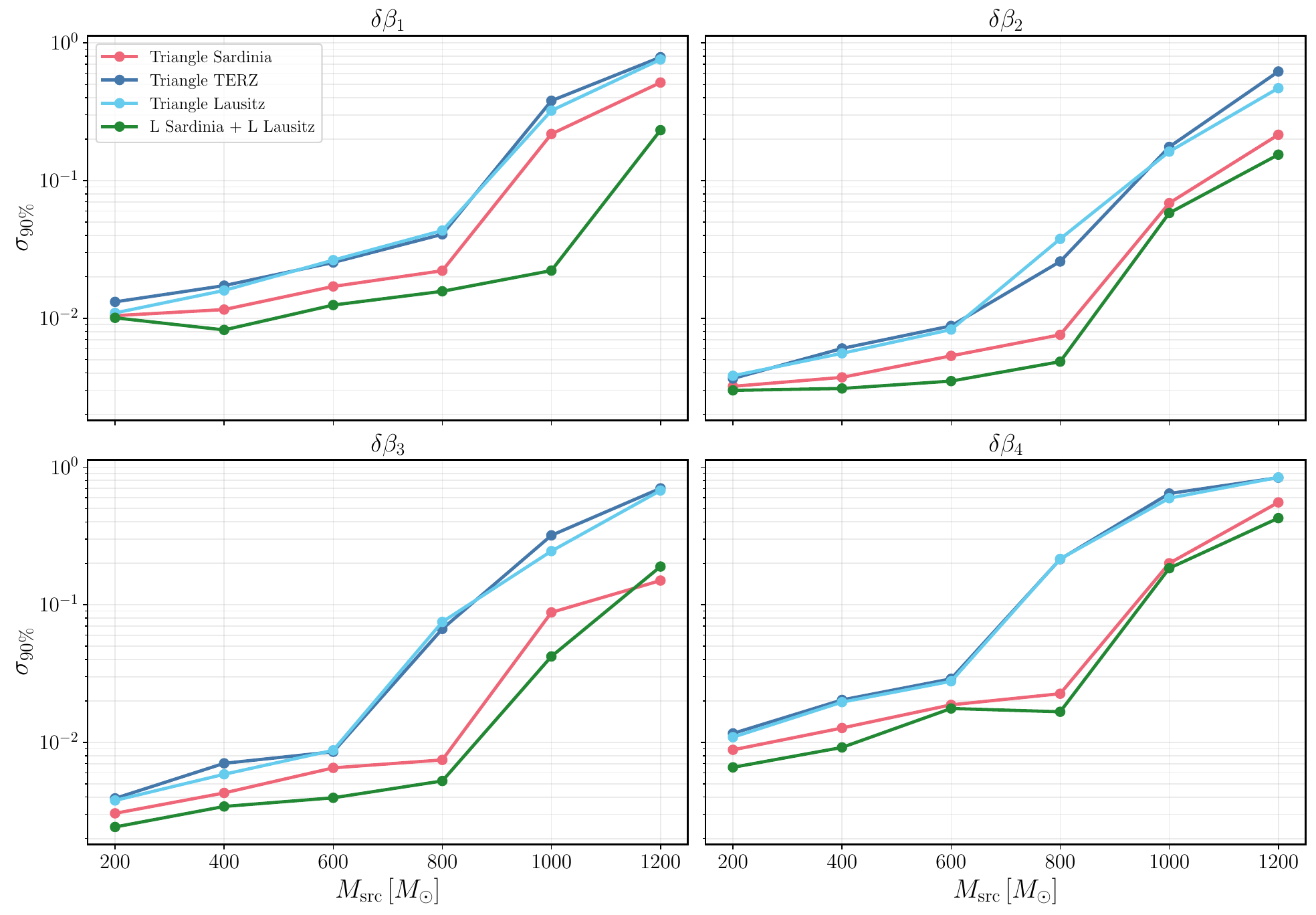}
    \caption{
    Half-width of the central $90\%$ credible interval, $\sigma_{90\%}$, for the beyond-GR deviations in the intermediate regime as a function
    of the injected source-frame total mass. The injections are performed at
    \(z=4\) and \(\iota=0.2\), and the different curves correspond to
    the detector configurations considered in this work.}
    \label{fig:bgr_intermediate}
\end{figure}

The intermediate regime describes the transition between the PN-dominated inspiral and the merger-ringdown portions of the signal. In the IMRPhenomX families, the phase in this regime is modeled as~\cite{Roy:2025gzv}
\begin{equation}
\Psi_{\rm Int}(f)
=
\frac{1}{\eta}
\left[
b_0'
+b_0 f
+b_1 \log f
-b_2 f^{-1}
-\frac{b_3}{2}f^{-2}
-\frac{b_4}{3}f^{-3}
-\frac{2c_L}{f_{\rm damp}}
\tan^{-1}\left(
\frac{f-f_{\rm RD}}{2f_{\rm damp}}
\right)
\right],
\end{equation}
where $f_{\rm RD}$ and $f_{\rm damp}$ denote the ringdown and damping frequencies, respectively. The coefficients $b_i$ and $c_L$ are phenomenological coefficients calibrated to numerical-relativity simulations. In the TIGER parameterization, deviations from GR are introduced through
\begin{equation}
b_i \rightarrow b_i^{\rm GR}(1+\delta b_i),
\qquad i=1,\ldots,4,
\end{equation}
so that the intermediate-regime deviation parameters are
\begin{equation}
\left\{
\delta b_1,\delta b_2,\delta b_3,\delta b_4
\right\}.
\end{equation}
As in the inspiral analysis, we vary one deviation parameter at a time in the recovery, while fixing all the remaining beyond-GR coefficients to zero. The resulting constraints are shown in Fig.~\ref{fig:bgr_intermediate}.

These coefficients primarily affect the portion of the waveform connecting the inspiral to the late coalescence. Their mass dependence therefore reflects the competition between two effects. As the total mass increases, the intermediate and post-inspiral portions of the signal contribute more prominently to the observed SNR. At sufficiently large masses, however, the corresponding characteristic frequencies shift toward the low-frequency edge of the detector band, where the sensitivity deteriorates.
The resulting mass dependence can differ among the individual deviation parameters and detector configurations.

\subsubsection{Ringdown}
\label{sec:parametrized_ringdown}
The merger-ringdown regime probes the late-time, highly dynamical stage of the coalescence. In the IMRPhenomX families, the phase in this regime is modeled as~\cite{Roy:2025gzv}
\begin{equation}
\Psi_{\rm MR}(f)
=
\frac{1}{\eta}
\left[
c_0'
+c_0 f
+\frac{3}{2}c_1 f^{2/3}
-c_2 f^{-1}
-\frac{c_4}{3}f^{-3}
+\frac{c_L}{f_{\rm damp}}
\tan^{-1}\left(
\frac{f-f_{\rm RD}}{f_{\rm damp}}
\right)
\right],
\end{equation}
where $\eta$ is the symmetric mass ratio, while $f_{\rm RD}$ and $f_{\rm damp}$ denote the ringdown and damping frequencies, respectively. The coefficients $c_i$ are phenomenological coefficients calibrated to numerical-relativity simulations. Within the TIGER framework, deviations from GR are introduced according to
\begin{equation}
c_i \rightarrow c_i^{\rm GR} (1+\delta{c}_i),
\end{equation}
for the coefficients entering the parametrized test, so that the merger-ringdown deviation parameters are
\begin{equation}
\left\{
\delta{c}_1,
\delta{c}_2,
\delta{c}_4,
\delta{c}_\ell
\right\}.
\end{equation}
These coefficients modify the merger-ringdown portion of the waveform and are therefore particularly relevant for high-mass systems, for which fewer inspiral cycles remain in band and an increasingly large fraction of the observed signal is contributed by the merger and ringdown.

\begin{figure}
    \centering
    \includegraphics[width=0.9\linewidth]{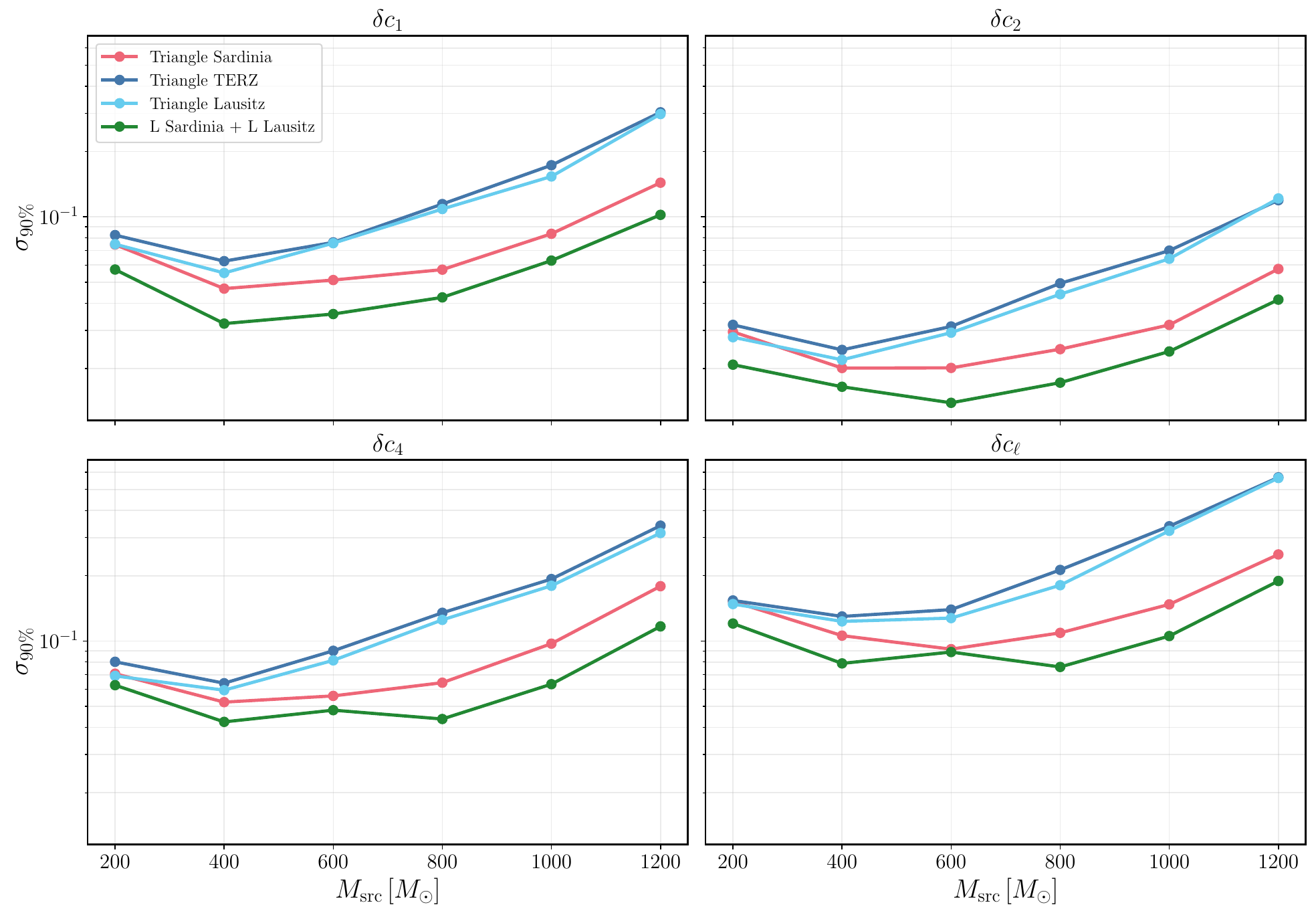}
    \caption{
    Half-width of the central $90\%$ credible interval, $\sigma_{90\%}$, for the beyond-GR deviations in the merger-ringdown regime as a function of the injected source-frame total mass. The injections are performed at
    $z=4$ and $\iota=0.2$, and the different curves correspond to
    the detector configurations considered in this work.
}
    \label{fig:bgr_ringdown}
\end{figure}

As in the previous analyses, we vary one deviation parameter at a time while fixing all the others to their GR values. The resulting constraints are shown in Fig.~\ref{fig:bgr_ringdown}.

The behaviour of $\sigma_{90\%}$ in this sector quantifies the ability of each detector configuration to constrain departures from GR in the strong-field regime, with smaller values corresponding to tighter bounds. As the total mass increases, the merger-ringdown contribution initially becomes more prominent and the constraints can improve. At sufficiently large masses, however, the characteristic frequencies shift toward the low-frequency edge of the detector band, where the sensitivity deteriorates, leading to a degradation of the bounds.

\subsection{Black hole spectroscopy}
\label{sec:bh_spectroscopy}
We consider a standard BH spectroscopy test
(see Ref.~\cite{Berti:2025hly} for a recent review of the topic), whose goal is to assess whether the frequencies and damping times of multiple QNMs are consistent with
those of a remnant Kerr BH. The ringdown signal is modeled as a superposition of
damped sinusoids,
\begin{align}
\label{eq:rdmodelFINAL}
h(t)
=
\sum_i A_i
\cos\left[
2\pi f_i^{\rm Kerr}
\left(1+\delta f_i\right)t+\phi_i
\right]
\exp\left[
-\frac{t}
{\tau_i^{\rm Kerr}\left(1+\delta\tau_i\right)}
\right].
\end{align}
Here, $A_i$ and $\phi_i$ are respectively the amplitudes and phases, while $f_i^{\rm Kerr}$ and $\tau_i^{\rm Kerr}$ are the GR frequency
and damping time of the $i$-th Kerr QNM, which are solely functions of $M_f$ and $\chi_f$, denoting respectively the remnant mass and dimensionless spin. The parameters $\delta f_i$ and
$\delta\tau_i$ describe fractional deviations from the corresponding Kerr
values. As in the TIGER framework, all deviation parameters vanish when GR is considered.
Following the ringdown analyses of Refs.~\cite{Capano:2021etf,Forteza:2022tgq}, we inject a signal composed of the fundamental $(2,2,0)$ and $(3,3,0)$ QNMs. The corresponding amplitudes and phases are obtained from the fitting formulas implemented in the \texttt{jaxqualin} pipeline~\cite{Cheung:2023vki}, evaluated at the progenitor parameters $(\eta,\chi_1,\chi_2)$. The injected values are listed in Table~\ref{tab:rd_injected_parameters}. We refer the reader to Ref.~\cite{Crescimbeni:2025ytx} for a detailed assessment of the accuracy of these amplitude and phase fits across the binary parameter space. The extrinsic parameters are fixed to the values reported in Table~\ref{tab:IMBH_injected_extrinsic}.

\begin{table}[t]
\centering
\begin{tabular}{lc}
\toprule
Parameter & Injected value \\
\midrule
$A_{220}$    & $1.0474101275$ \\
$\phi_{220}$ & $0$ \\
$A_{330}$    & $0.0940858294$ \\
$\phi_{330}$ & $0.4103742007$ \\
\bottomrule
\end{tabular}
\caption{Injected amplitudes and phases of the $(2,2,0)$ and $(3,3,0)$ QNMs used in the ringdown analysis. The values are obtained from the fitting formulas implemented in the \texttt{jaxqualin} pipeline~\cite{Cheung:2023vki}.}
\label{tab:rd_injected_parameters}
\end{table}

In the recovery, we sample over the remnant mass and spin, the amplitudes and phases of the two modes, and the fractional deviations of the subdominant-mode frequency and damping time:
\begin{equation}
\vec{\theta}_{\rm RD}
=
\left\{
M_f,\chi_f,
A_{220},\phi_{220},
A_{R,330},\phi_{R,330},
\delta f_{330},\delta\tau_{330}
\right\},
\end{equation}
where
\begin{equation}
\begin{cases}
A_{R,330}=\frac{A_{330}}{A_{220}}\\
\phi_{R,330}=\frac{3}{2}\phi_{220}-\phi_{330}.    
\end{cases}
\end{equation}
For simplicity, we fix the sky location to its injected value, since its impact on the ringdown inference is expected to be marginal for the present analysis~\cite{Finch:2022ynt,Correia:2023ipz,Dey:2026btx}. We also fix the inclination angle, which is strongly correlated with the mode amplitudes. Finally, we do not vary $\delta f_{220}$ or $\delta\tau_{220}$, since deviations in the dominant-mode frequency and damping time are degenerate with the remnant mass and spin. The $(2,2,0)$ mode is therefore used to infer the Kerr parameters $(M_f,\chi_f)$, while the $(3,3,0)$ mode is allowed to depart from the corresponding Kerr prediction.

We perform the analysis for the same four detector configurations considered above using \texttt{PyCBC Inference}~\cite{Biwer:2018osg}. The relative constraining power of the different configurations is quantified through the posterior widths of $\delta f_{330}$ and $\delta\tau_{330}$. Since the injected signal is consistent with GR, both parameters are expected to be compatible with zero. Smaller credible intervals therefore correspond to a more precise spectroscopy measurement and to a stronger consistency test of the Kerr nature of the remnant.

\begin{figure}
    \centering
    \includegraphics[width=0.46\linewidth]{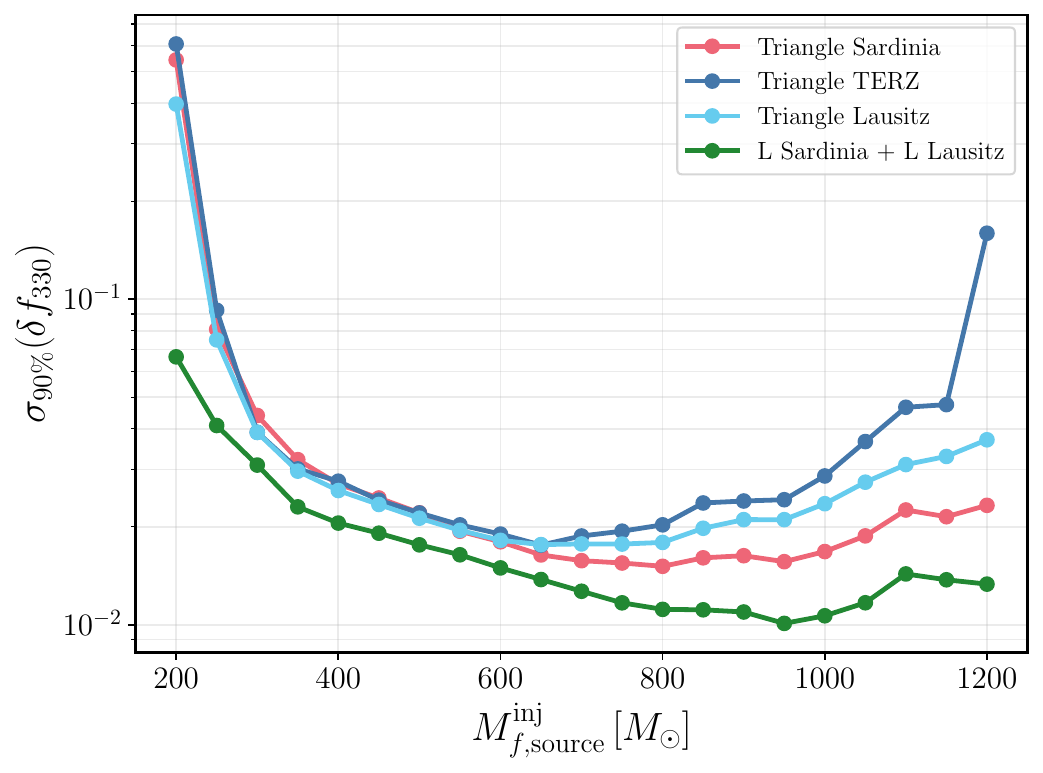}
    \includegraphics[width=0.46\linewidth]{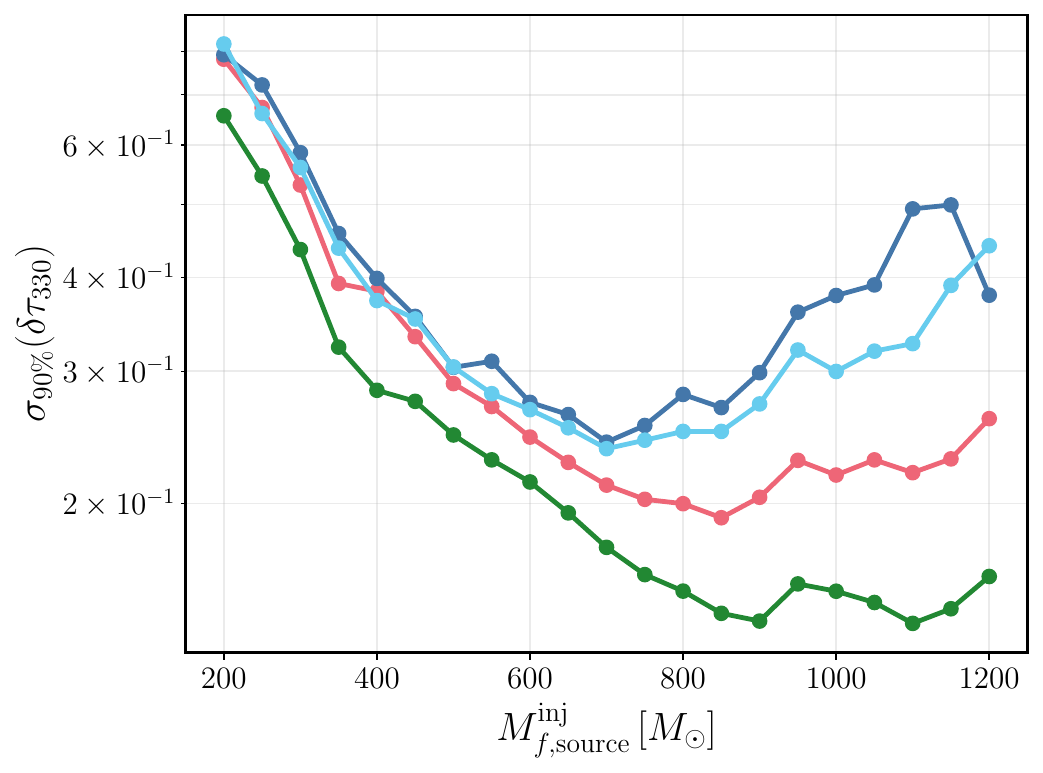}
    \caption{Symmetric $90\%$ credible half-width of the fractional ringdown deviations, $\delta f_{330}$ and $\delta \tau_{330}$, as functions of the injected remnant mass, for the different detector configurations.
    }
    \label{fig:ringdown}
\end{figure}

The results are shown in Fig.~\ref{fig:ringdown}. 
For both deviation parameters, the uncertainties initially decrease with increasing injected remnant mass, reach a broad minimum at
$M_{f,\rm source}^{\rm inj}\sim {\rm few}\times10^2\,M_\odot$,
with the precise location depending on the observable and detector configuration, and increase again toward the largest masses considered.
This nonmonotonic behaviour reflects the competition between the increasing ringdown SNR and the progressive shift of the QNM frequencies toward the low-frequency edge of the detector band, where the sensitivity deteriorates. As expected, the damping time is systematically more difficult to measure than the oscillation frequency, with the uncertainty on $\delta\tau_{330}$ being approximately one order of magnitude larger than that on $\delta f_{330}$.

At lower masses, the differences among the triangular configurations are modest because the $(3,3,0)$ mode lies in a frequency range where their sensitivities are comparable. The impact of the site-dependent noise becomes increasingly important at larger masses, when the ringdown signal moves below approximately $10\,{\rm Hz}$. In this regime, the Sardinia triangle yields substantially tighter constraints than the TERZ and Lausitz triangles, whose uncertainties increase more rapidly.

The 2L configuration provides the strongest constraints across the full mass range considered. Near the optimal mass range, it improves the measurement of both $\delta f_{330}$ and $\delta\tau_{330}$ by a factor of approximately $1.5$--$2$ relative to the Sardinia triangle, while the improvement with respect to the TERZ and Lausitz triangles becomes significantly larger at the highest masses. These results show that BH spectroscopy of intermediate-mass remnants is particularly sensitive to the achievable low-frequency noise level and that higher site noise can substantially offset the gain associated with longer detector arms. As in the previous analyses, however, the superior performance of the 2L network at high masses relies on the assumption that at least one of the two L-shaped detectors achieves a low-frequency sensitivity comparable to that of the Sardinia site.

\subsection{Binary neutron stars: early warnings and sky localizations}
Finally, we investigate the impact of the different low-frequency noise configurations on binary NS mergers, focusing on early-warning times and sky-localization capabilities. These are among the most direct connections between the low-frequency performance of ET and multimessenger astronomy: sensitivity at a few Hz determines how early a source can be identified, while the long signal duration, Earth-rotation modulation, and detector-network geometry determine how accurately it can be localized before merger~\cite{ET:2025xjr}. 

As a reference source, we consider a GW170817-like binary~\cite{LIGOScientific:2017vwq} at redshift $z\simeq0.0098$, corresponding to a luminosity distance $d_L\simeq44\,{\rm Mpc}$, and with inclination $\iota\simeq2.56\,{\rm rad}$ (see Table~\ref{tab:BNS_injected_parameters} for the full set of injected parameters). 
We neglect tidal corrections by setting $\Lambda_1=\Lambda_2=0$, since their contribution is relevant mainly during the late inspiral~\cite{Chakraborty:2026qru,Rodriguez:2026iot}. At the maximum frequencies reached by the signals truncated $3\,{\rm h}$,
$1\,{\rm h}$, and $30\,{\rm min}$ before merger, tidal contributions to the phase are negligible for the localization quantities considered here. We then generate the signal with a \texttt{TaylorF2} waveform~\cite{Damour:2000zb}, which faithfully describes the early inspiral of the signal.
\begin{table}[t]
\centering

\begin{subtable}{0.48\linewidth}
\centering
\begin{tabular}{cc}
\toprule
Intrinsic parameter & Injected value \\
\midrule
\(M_{\rm tot}^{\rm src}\) & \(2.8\,M_\odot\) \\
\(m_1^{\rm src}\) & \(1.6092\,M_\odot\) \\
\(m_2^{\rm src}\) & \(1.1908\,M_\odot\) \\
\(q\) & \(0.74\) \\
\(\chi_1\) & \(0\) \\
\(\chi_2\) & \(0\) \\
\(\Lambda_1\) & \(0\) \\
\(\Lambda_2\) & \(0\) \\
\bottomrule
\end{tabular}
\caption{Intrinsic parameters.}
\label{tab:BNS_injected_intrinsic}
\end{subtable}
\hfill
\begin{subtable}{0.48\linewidth}
\centering
\begin{tabular}{cc}
\toprule
Extrinsic parameter & Injected value \\
\midrule
\(z\) & \(0.0098\) \\
\(d_L\) & \(43.70\,{\rm Mpc}\) \\
\(\iota\) & \(146.68^\circ\) \\
\(\psi\) & \(89.00^\circ\) \\
\(\phi_c\) & \(1.3\,{\rm rad}\) \\
\(t_c\) & \(1187008882.43\,{\rm s}\) \\
\(\alpha\) & \(197.45^\circ\) \\
\(\delta\) & \(23.38^\circ\) \\
\bottomrule
\end{tabular}
\caption{Extrinsic parameters.}
\label{tab:BNS_injected_extrinsic}
\end{subtable}

\caption{
Injected parameters used in the Bayesian inference analysis of the
GW170817-like binary neutron-star source with \texttt{bilby-x-g}. The sky position corresponds to
NGC~4993. The same injection is employed for all detector
configurations and signal truncation times.
}
\label{tab:BNS_injected_parameters}
\end{table}
Because the signal remains in band for several hours, we account for the time dependence of the detector response induced by Earth rotation~\cite{Baral:2025geo}.

\begin{figure}[t]
    \centering
    \includegraphics[width=0.9\linewidth]{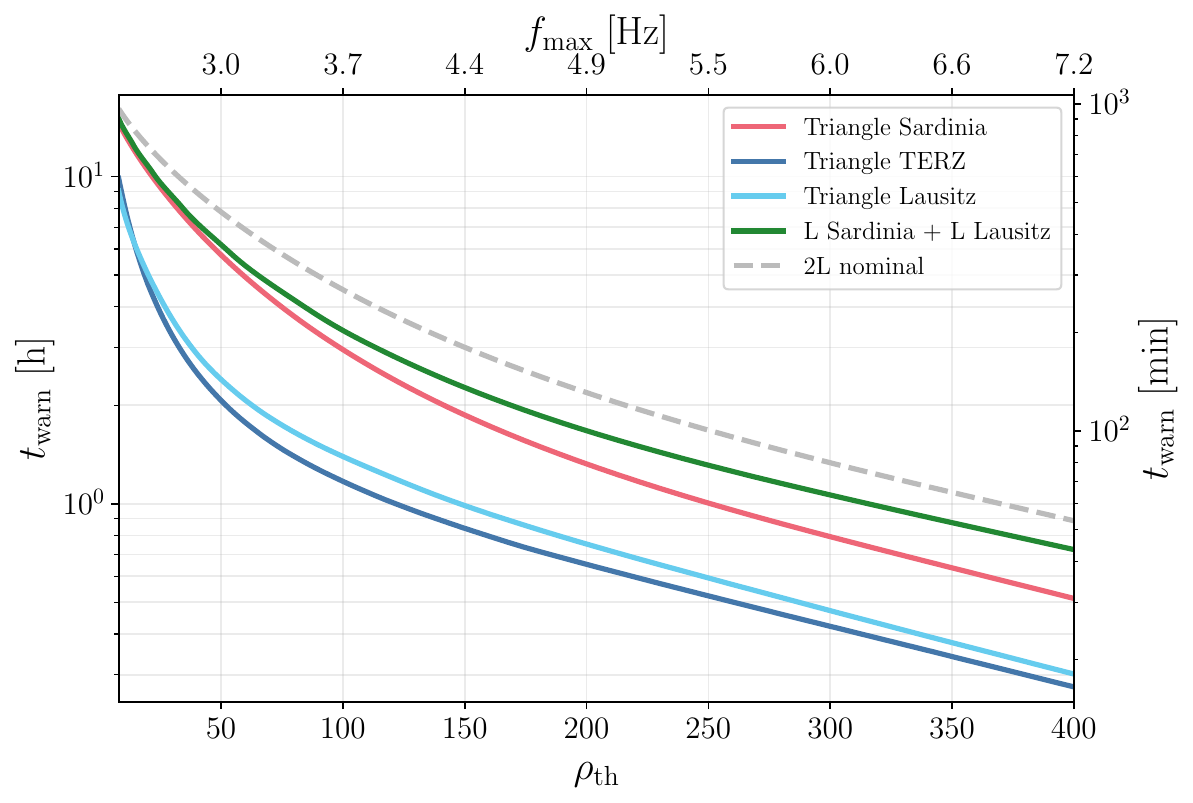}
    \caption{Warning time $t_{\rm warn}$ for a GW170817-like binary, averaged over sky position and polarization, as a function of the adopted network-SNR threshold $\rho_{\rm th}$. The different curves correspond to the detector configurations considered in this work, including the nominal 2L sensitivity. The upper horizontal axis shows the sky-averaged maximum frequency $f_{\rm max}$ at which the nominal 2L network crosses the corresponding SNR threshold.}  
    \label{fig:warning}
    
\end{figure}

We sample the angular parameters $(\alpha,\delta,\psi)$ on a $12\times12\times12$ grid. For each point, we compute the warning time $t_{\rm warn}(\alpha,\delta,\psi)$, defined as the time before merger at which the accumulated network SNR $\rho(\alpha,\delta,\psi)$ first reaches a chosen threshold $\rho_{\rm th}$. At leading PN order, this time is related to the corresponding upper integration frequency by
\begin{equation}
    t_{\rm warn}(\alpha,\delta,\psi)
    =
    \frac{5}{256}
    \left(
    \frac{G\mathcal{M}^{\rm det}}{c^3}
    \right)^{-5/3}
    \left[
    \pi f_{\rm max}(\alpha,\delta,\psi)
    \right]^{-8/3},
    \label{time_warn}
\end{equation}
where $f_{\rm max}(\alpha,\delta,\psi)$ is the GW frequency at which
\begin{equation}
    \rho\bigl(f_{\rm max};\alpha,\delta,\psi\bigr)
    =
    \rho_{\rm th}.
\end{equation}
We then average $t_{\rm warn}$ over $(\alpha,\cos\delta,\psi)$, ensuring a uniform sampling of the sky and polarization, and repeat the calculation for all detector networks introduced in Sec.~\ref{sec:detectors}. We checked that the adopted $12^3$ grid is sufficient for convergence of the average. The results are shown in Fig.~\ref{fig:warning} as a function of $\rho_{\rm th}$.

As expected, the warning time decreases as the SNR threshold increases, since a larger portion of the inspiral must be accumulated before the event is identified. The assumed low-frequency noise has a substantial impact on this quantity. For $\rho_{\rm th}=50$, the Sardinia triangle and the site-dependent 2L network yield mean warning times of approximately $5\,{\rm h}$, roughly twice the ${\sim}2\,{\rm h}$ obtained for the TERZ and Lausitz triangles. The difference remains significant at higher thresholds: for $\rho_{\rm th}=250$, the Sardinia triangle and the site-dependent 2L network still provide approximately $1\,{\rm h}$ of warning, compared with ${\sim}30\,{\rm min}$ for the other triangular configurations.

The nominal 2L network provides the longest warning time over the full range of thresholds considered.
Comparing it with the site-dependent 2L configuration shows that the advantage expected from the longer arms can be substantially offset by diminished low-frequency sensitivity. Among the site-dependent networks, the 2L configuration provides the longest warning time over the full range of thresholds considered, although at lower thresholds the Sardinia triangle performs nearly as well. The difference between the two becomes more pronounced at higher thresholds, reflecting the combined effects of detector sensitivity and antenna response after averaging over source position and polarization.

The gain is operationally important: several additional hours would provide substantially more time to validate an alert, schedule follow-up observations, and point electromagnetic facilities before the merger. Thus, for  binary NSs, the scientific return of improved sensitivity at a few Hz is not captured only by the final SNR or parameter uncertainties, but also by when the source first becomes observable.

\begin{figure*}[htbp]
    \centering

    \begin{minipage}[c]{0.42\textwidth}
        \centering
        \includegraphics[width=\textwidth]{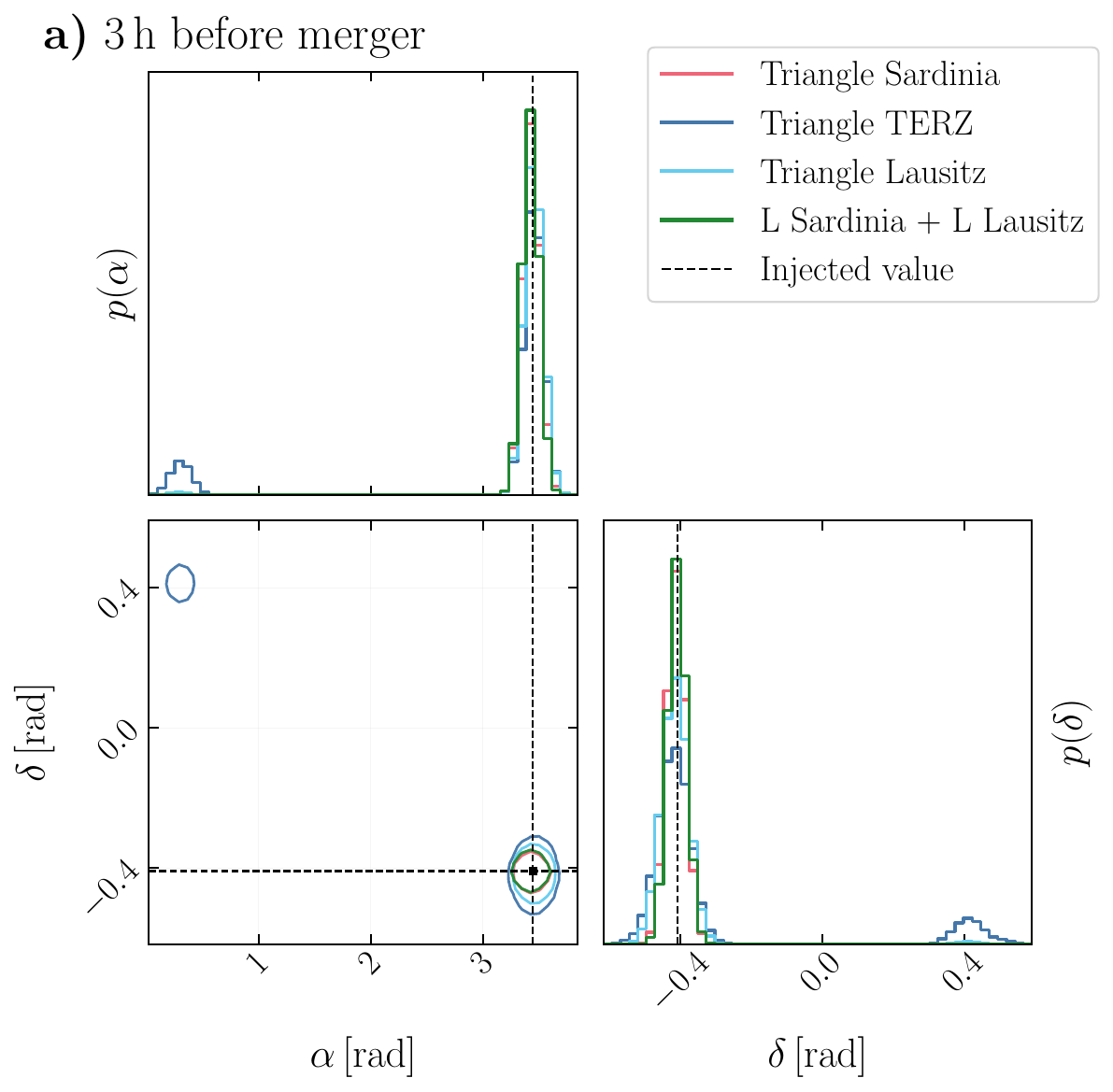}
    \end{minipage}
    \hfill
    \begin{minipage}[c]{0.40\textwidth}
        \centering
        \includegraphics[width=\textwidth]{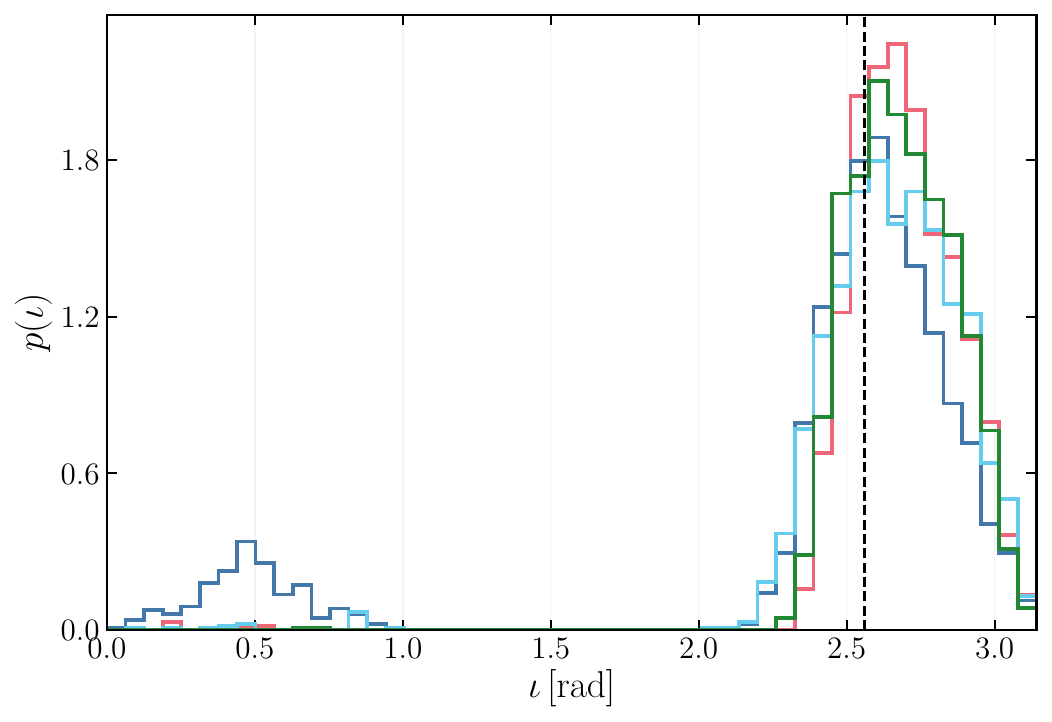}
    \end{minipage}

    \vspace{0.25cm}

    \begin{minipage}[c]{0.42\textwidth}
        \centering
        \includegraphics[width=\textwidth]{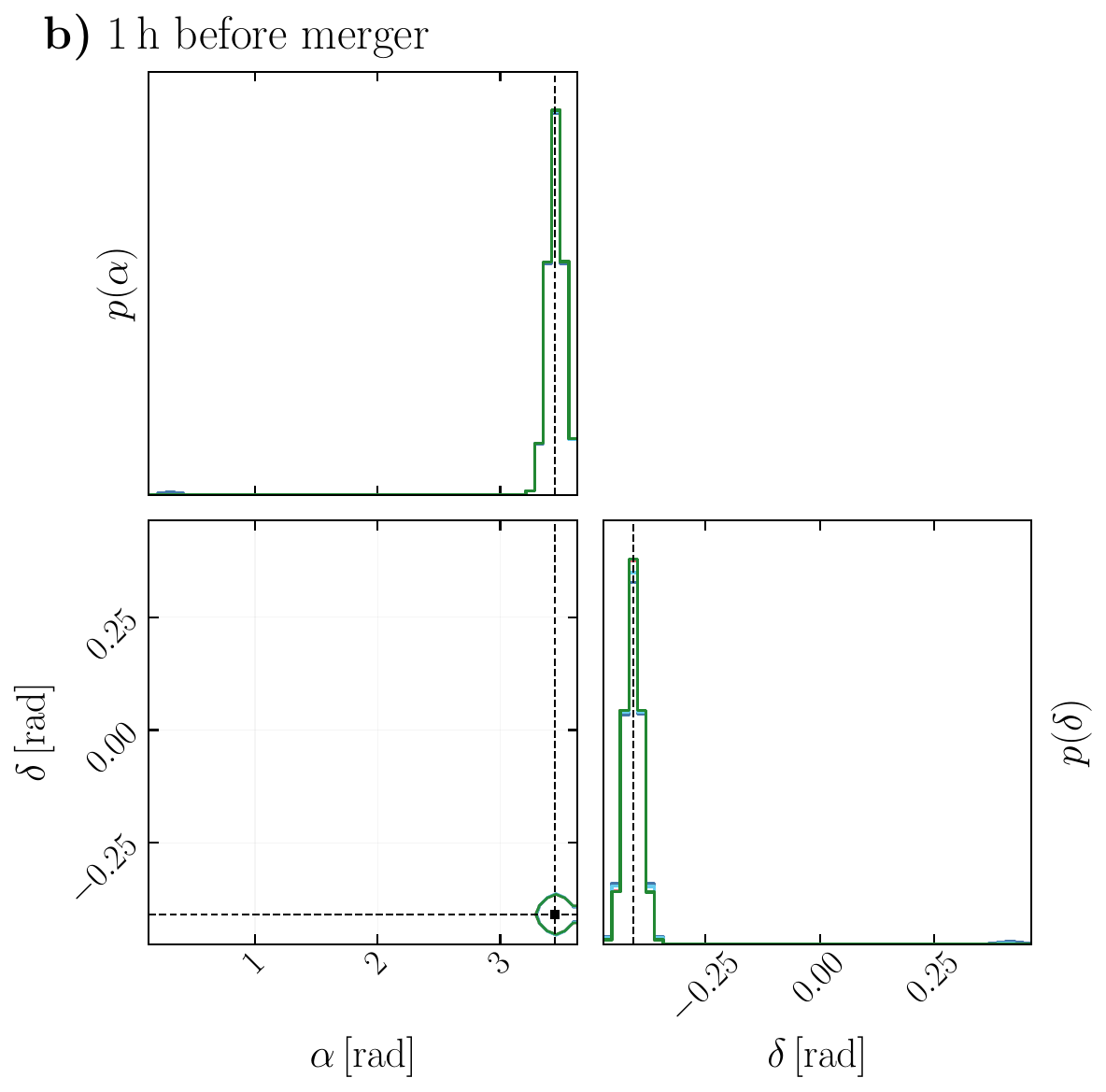}
    \end{minipage}
    \hfill
    \begin{minipage}[c]{0.40\textwidth}
        \centering
        \includegraphics[width=\textwidth]{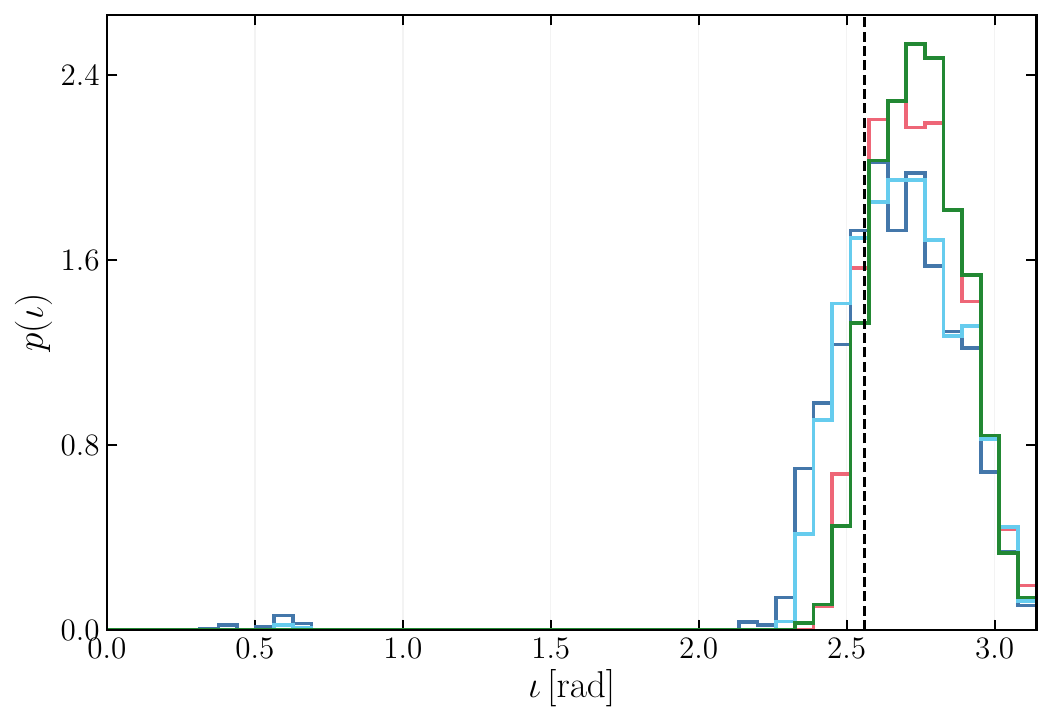}
    \end{minipage}

    \vspace{0.25cm}

    \begin{minipage}[c]{0.42\textwidth}
        \centering
        \includegraphics[width=\textwidth]{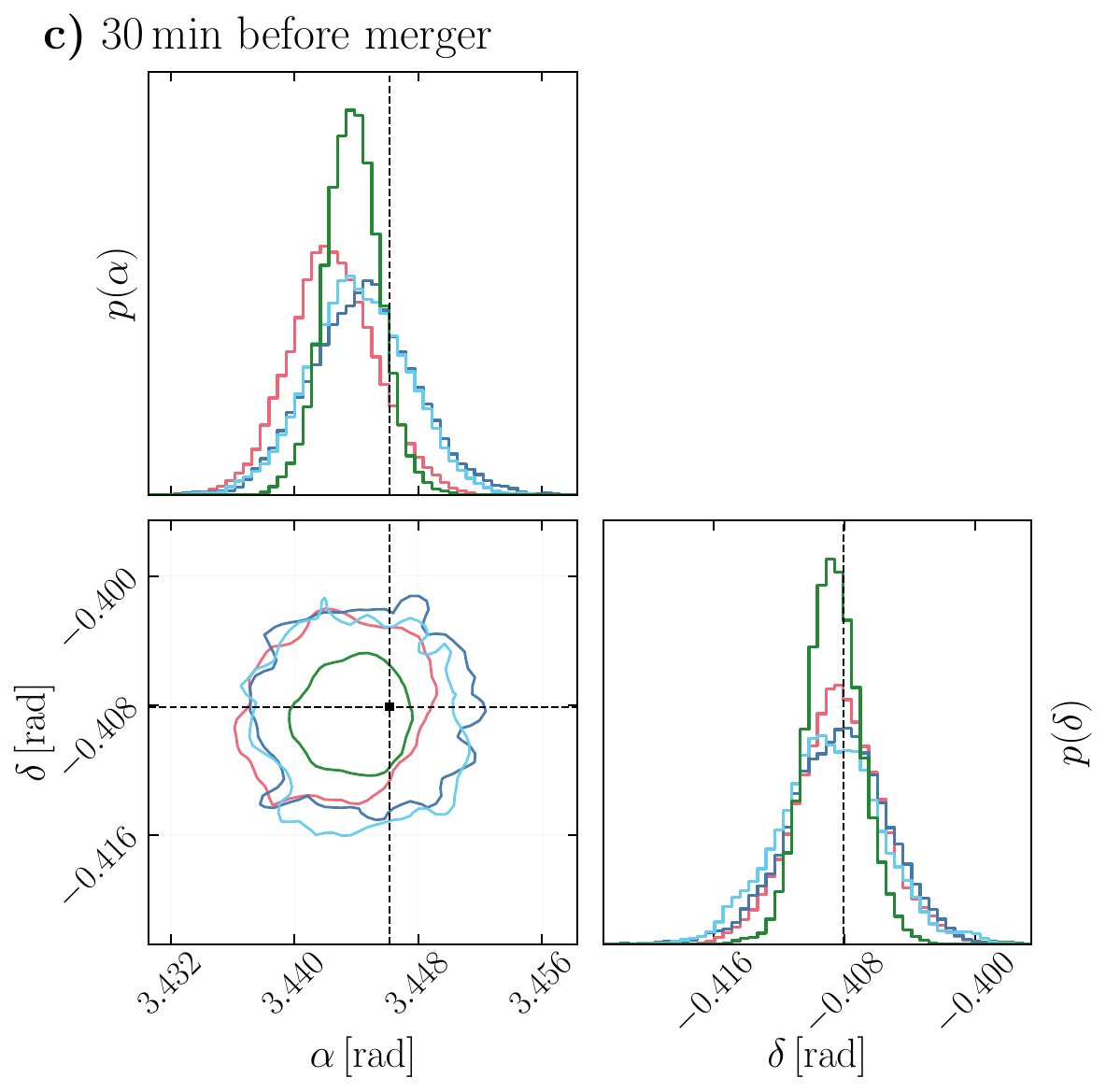}
    \end{minipage}
    \hfill
    \begin{minipage}[c]{0.40\textwidth}
        \centering
        \includegraphics[width=\textwidth]{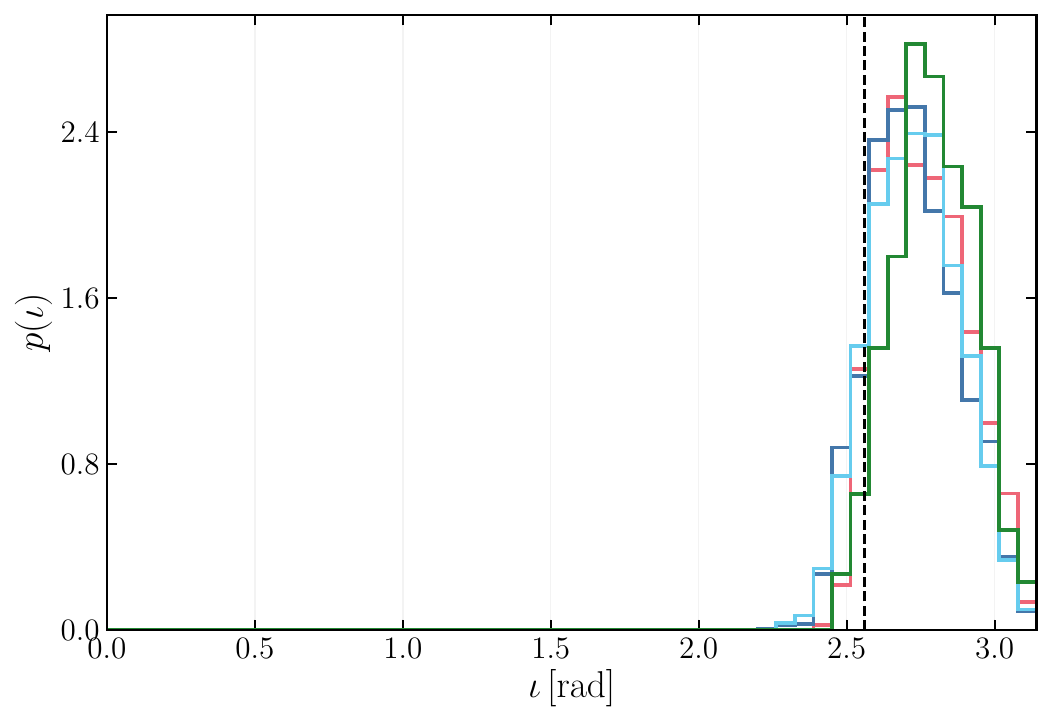}
    \end{minipage}

    \caption{Posterior distributions of the sky localization parameters for a GW170817-like event obtained using the full prior range of $\iota$. Panels a), b), and c) correspond to $T=10800\,\mathrm{s}$, $T=3600\,\mathrm{s}$, and $T=1800\,\mathrm{s}$ before merger, respectively. In each panel, the left plot shows the joint and marginalized posterior distributions of right ascension $\alpha$ and declination $\delta$, while the right plot shows the one-dimensional posterior probability density $p(\iota)$ of the inclination angle. The corresponding plot obtained by enforcing the prior $\iota\in[\pi/2,\pi]$, hence removing multimodalities, is presented in the Appendix, see Fig.~\ref{fig:posteriors_BNS_theta_geq_pi2}.}
    \label{fig:posteriors_BNS_standard}
\end{figure*}

We next quantify how accurately the source can be localized while it is still a sufficient time away from merger. We perform parameter estimation for a GW170817-like event, with an injected sky location of Ref.~\cite{LIGOScientific:2017vwq} \footnote{Since the localization
analysis is based on a single GW170817-like source, the absolute areas and their
ordering should not be interpreted as population-averaged forecasts; these
will depend on the source position, orientation, distance, and alert time.} (see Table~\ref{tab:BNS_injected_parameters}), using \texttt{bilby-x-g}~\cite{BaralBilbyXG}, which adapts the current inference pipeline to third-generation detectors by including Earth rotation. The full prior ranges are reported in the Appendix (see Table~\ref{tab:BNS_priors}). 
For each configuration, the analysis includes only the signal accumulated up to $t_{\rm warn}=3\,{\rm h}$, $t_{\rm warn}=1\,{\rm h}$ and $t_{\rm warn}=30\,{\rm min}$ before merger. For each case, we compute the $90\%$ solid angle value, starting from the posterior samples of $\alpha$ and $\delta$, as follows. Defining $u=\sin\delta$, for which
\begin{equation}
\mathrm{d}\Omega
=
\mathrm{d}u\,\mathrm{d}\alpha,
\end{equation}
the posterior density is reconstructed on a uniform grid in the equal-area coordinates $(\alpha,u)$ using a periodic kernel-density estimator in $\alpha$, so that each grid cell subtends the same solid angle,
\begin{equation}
\Delta\Omega_{\rm cell}
=
\Delta\alpha\,\Delta u.
\end{equation}
The cells are then ranked by posterior probability, and the $90\%$ credible region is defined as the smallest set containing at least $90\%$ of the total probability. Its area is therefore
\begin{equation}
\Delta\Omega_{90\%}
=
N_{90\%}\Delta\Omega_{\rm cell}.
\end{equation}
The corresponding values of $\Delta\Omega_{90\%}$ are reported in Table~\ref{tab:solid_angles}, while the posterior distributions for $(\alpha,\delta,\iota)$ are shown in Fig.~\ref{fig:posteriors_BNS_standard}. We verified that the reported areas are stable under variations of the KDE bandwidth and grid resolution.

\begin{table}[htbp]
    \centering
    \begin{tabular}{lccc}
        \hline
        Configuration
        & \multicolumn{3}{c}{$\Delta\Omega_{90\%}\,[\mathrm{deg}^2]$} \\
        \cline{2-4}
        & $3\,{\rm h}$
        & $1\,{\rm h}$
        & $30\,{\rm min}$ \\
        \hline
        Triangle TERZ            & $118.3$ & $2.29$ & $0.49$ \\
        Triangle Lausitz         & $44.4$ & $1.59$ & $0.48$ \\
        Triangle Sardinia        & $6.09$ & $0.60$ & $0.37$ \\
        L Sardinia + L Lausitz   & $3.94$ & $0.32$ & $0.16$ \\
        \hline
    \end{tabular}
    \caption{Sky-localization area enclosing $90\%$ of the posterior probability for the four ET configurations considered in this work, using the signal accumulated up to $3\,{\rm h}$, $1\,{\rm h}$ and $30\,{\rm min}$ before merger, for a GW170817-like source.
    These values refer to a single injection and should not be interpreted as population-averaged localization forecasts. The corresponding results obtained by enforcing the prior $\iota\in[\pi/2,\pi]$, hence removing multimodalities, are presented in the Appendix, see Table~\ref{tab:solid_angles_theta}.   }
    \label{tab:solid_angles}
\end{table}

For the analysis truncated three hours before merger, the 2L configuration provides the tightest localization, followed by the Sardinia triangle. In particular, the 2L network yields
$\Delta\Omega_{90\%}=3.94\,{\rm deg}^2$, while the Sardinia triangle gives
$\Delta\Omega_{90\%}=6.09\,{\rm deg}^2$. The localization areas obtained with the Lausitz and TERZ triangles are instead significantly larger than the 2L result, by factors of approximately $11$ and $30$, respectively. This degradation reflects both the higher low-frequency noise of the adopted TERZ and Lausitz realizations and the presence of multimodal structures in the $(\alpha,\delta)$ posterior, as shown in panel a) of Fig.~\ref{fig:posteriors_BNS_standard}.


These multimodalities are associated with the inclination degeneracy
$\iota\rightarrow\pi-\iota$~\cite{Usman:2018imj, Mascioli:2025cnw, Santoliquido:2025aiq}. At leading quadrupolar order, the plus-polarization amplitude is unchanged under this transformation, while the cross-polarization amplitude changes sign. This can be approximately compensated by corresponding transformations of the polarization and sky-position parameters. As a result, two distinct modes appear in the $(\alpha,\delta)$ posterior, one of which lies close to the antipodal point of the injected sky position. To assess the impact of this degeneracy, we repeated the analysis by restricting the inclination prior to $\iota\in[\pi/2,\pi]$. 
In this case, the localization area for the TERZ triangle is reduced by approximately a factor of $2$, whereas the Lausitz result remains essentially unchanged
(see Table~\ref{tab:solid_angles_theta} in the Appendix).

However, even after forcing the absence of the multimodalities by restricting the inclination prior, the localization areas
of the Lausitz and TERZ triangles $3\,{\rm h}$ before merger remain approximately one order of magnitude larger than those obtained with the Sardinia triangle and the 2L configuration.

One hour before merger, the 2L network again provides the tightest localization, with
$\Delta\Omega_{90\%}=0.32\,{\rm deg}^2$. This area is smaller by factors of approximately $2$, $5$, and $7$ relative to the Sardinia, Lausitz, and TERZ triangles, respectively. The Sardinia triangle nevertheless achieves sub-square-degree localization,
$\Delta\Omega_{90\%}=0.60\,{\rm deg}^2$, improving over the Lausitz and TERZ triangles by factors of approximately $2.5$ and $4$, respectively.
As shown in panel b) of Fig.~\ref{fig:posteriors_BNS_standard}, the posteriors are essentially unimodal at this stage. For the TERZ and Lausitz triangles, however, the preferred mode lies close to the antipodal sky position, whereas the other configurations favor the region containing the injected location. We stress that this bias should not be interpreted as an intrinsic limitation of the TERZ and Lausitz configurations. Rather, it is likely contingent on the particular sky position and orientation of GW170817, which determine how the nearly degenerate sky modes are weighted by the antenna response of each detector. For a different source location, an analogous mode selection could instead occur for one of the other triangular configurations. The result therefore illustrates that the resolution of these degeneracies depends not only on the low-frequency sensitivity, but also on the detector location, orientation, and source position. In general, the geographically separated 2L network provides the most robust reconstruction, owing to its timing baseline and antenna-pattern diversity. A systematic investigation over a broader source population, including the possible role of precession and higher-order modes, is left for future work; see Ref.~\cite{Mascioli:2025cnw} for a recent analysis in this direction.

At $30\,{\rm min}$ before merger, as shown in panel c) of Fig. \ref{fig:posteriors_BNS_standard}, all configurations achieve sub-square-degree localization. The 2L network remains the best-performing configuration, with $\Delta\Omega_{90\%}=0.16\,{\rm deg}^2$, followed by the Sardinia triangle with $\Delta\Omega_{90\%}=0.37\,{\rm deg}^2$. The TERZ and Lausitz triangles yield comparable localization areas, $\Delta\Omega_{90\%}=0.49\,{\rm deg}^2$ and $\Delta\Omega_{90\%}=0.48\,{\rm deg}^2$, respectively. At this stage, the localization area of the 2L network is smaller than those of the triangular configurations by factors of approximately $2.3$--$3$, while the differences among the three triangular networks are considerably less pronounced.

Together with Fig.~\ref{fig:warning}, these results highlight two complementary ingredients of binary-NS multimessenger performance. Good sensitivity at a few Hz allows the source to cross the detection threshold several hours earlier and extends the time baseline over which Earth rotation modulates the detector response. The geographically separated 2L network additionally benefits from timing information and antenna-pattern diversity, which account for its stronger localization for this particular injection. At the same time, the ordering among the three triangular configurations does not simply follow the corresponding noise curves, confirming that detector location and orientation also contribute nontrivially to sky localization.

\section{Conclusions}
\label{sec:conclusions}
In this work, we investigated how representative site-dependent noise configurations affect a selected set of ET science cases that rely particularly strongly on the detector sensitivity below approximately $10\,{\rm Hz}$. We compared triangular configurations with $10\,{\rm km}$ arms at the Lausitz, Sardinia, and TERZ sites with a 2L network composed of two geographically separated L-shaped detectors with $15\,{\rm km}$ arms located in Sardinia and Lausitz. We stress that the adopted curves should be interpreted as representative sensitivity scenarios based on the currently available site-characterization studies, rather than as definitive predictions of the final performance attainable at any candidate site.

Our results consistently show that the achievable low-frequency sensitivity can have a significant impact on ET science. For relatively light binary BHs, whose observable signal lies mainly at frequencies where the different noise curves are similar, the triangular configurations yield comparable performance. As the source mass or redshift increases, however, a progressively larger fraction of the signal moves below approximately $10\,{\rm Hz}$. In this regime, the improved sensitivity associated with the Sardinia noise configuration leads to larger SNRs, a greater redshift horizon, and more accurate measurements of the source parameters than the TERZ and Lausitz configurations.

For intermediate-mass BH binaries with source-frame masses of a few $\times10^3\,M_\odot$, the Sardinia triangle reduces the uncertainties on the detector-frame chirp mass and effective spin by factors of approximately $1.5$--$3$, depending on the parameter and source mass, relative to the other triangular configurations.

A similar hierarchy emerges in parametrized tests of GR. The largest differences occur for deviations whose constraining power is concentrated in the earliest inspiral, especially negative-PN corrections, or in the merger and ringdown of sufficiently massive systems. These are precisely the cases for which the relevant waveform information is most strongly weighted toward the low-frequency part of the ET band.

The BH-spectroscopy analysis leads to the same qualitative conclusion. For sufficiently massive remnants, the frequencies of the dominant and subdominant QNMs shift toward the low-frequency edge of the detector band, where the site-dependent noise becomes increasingly important. In this regime, the Sardinia triangle provides substantially tighter constraints on the frequency and damping time of the subdominant mode than the TERZ and Lausitz triangles. The damping time remains more difficult to measure than the oscillation frequency, but the relative ordering among the detector configurations is consistent for both observables.

Binary NSs provide a complementary illustration of the importance of the low-frequency band. Although the merger occurs at much higher frequencies, sensitivity at a few Hz determines how early the inspiral can be identified and how long Earth rotation can modulate the detector response before merger. For a GW170817-like source, the Sardinia triangle and the site-dependent 2L network provide warning times of several hours, approximately twice those obtained for the TERZ and Lausitz triangles over a broad range of detection thresholds. This gain is operationally important, as it directly increases the time available for alert validation and coordinated electromagnetic follow-up. The localization results show that low-frequency sensitivity and network geometry play complementary roles. 
One hour before merger, the 2L network provides the tightest localization, with $\Delta\Omega_{90\%}=0.32\,{\rm deg}^2$. This area is smaller by factors of approximately $2$, $5$, and $7$ relative to the Sardinia, Lausitz, and TERZ triangles, respectively. The Sardinia triangle nevertheless achieves a localization area below one square
degree and improves over the Lausitz and TERZ triangles by factors of approximately $2.5$ and $4$, respectively.

The differences among the triangular layouts are not determined by the noise curves alone, confirming that detector location and orientation also contribute nontrivially to sky localization. Since this analysis is based on a single GW170817-like injection, the absolute areas and their ordering should not be interpreted as population-averaged forecasts. Among the configurations considered here, the 2L network generally provides the best performance, as expected from its longer arms and geographical separation~\cite{Branchesi:2023mws}. This conclusion is nevertheless conditional on the assumed low-frequency noise. In the scenarios studied here, the advantage of the 2L network for massive systems relies on one of its two interferometers achieving a sensitivity comparable to that assumed for the Sardinia site. If both detectors were affected by a higher noise budget below approximately $10\,{\rm Hz}$, the gain associated with increasing the arm length from $10\,{\rm km}$ to $15\,{\rm km}$ would be significantly reduced and, for some observables, could be largely offset. Conversely, in some high-mass regimes the Sardinia triangle approaches the performance of the 2L configuration, showing that the achievable low-frequency sensitivity can be at least as important as the arm length itself.

An additional caveat concerns the recently identified projection of vertical
suspension thermal noise into the longitudinal interferometer direction
~\cite{ETVerticalThermalNoise1,ETVerticalThermalNoise2}, which is not included
in the sensitivity curves used here. Although this contribution is not site-dependent, it might be significant and, if not sufficiently mitigated, it could modify the relative low-frequency
performance of the $10\,{\rm km}$ and $15\,{\rm km}$ configurations. A similar mitigation strategy should also be adopted for site-dependent magnetic noise, which is currently being characterized at all candidate sites.
This reinforces the main conclusion of our study:
comparisons between detector geometries and arm lengths must ultimately be
performed using complete and realistic low-frequency noise budgets.

Our analysis assumes stationary Gaussian noise, statistically independent detector channels, and full network operation. These assumptions are useful for isolating the impact of the site-dependent sensitivity curves, but they neglect several effects that may alter the relative performance of the two geometries. A particularly important issue for the triangular configuration is the possible presence of correlated environmental noise between its co-located interferometers. Seismic, Newtonian, and magnetic disturbances may generate correlated noise and coincident glitches, invalidate the assumption of statistically independent channels, limit stochastic-background searches, and complicate the use of the null stream for detector characterization and glitch mitigation~\cite{Harms:2026ptf}. The extent of these effects will depend on the actual environmental couplings and on the effectiveness of magnetic shielding, Newtonian noise cancellation, and other mitigation strategies. 
Since such correlations are not included here, they could reduce or complicate some of the advantages of the triangular layout, particularly for science cases and detector-characterization methods that rely on independent co-located channels.

The impact of detector availability may also depend strongly on the operational and maintenance model~\cite{Negri:2026gwp}. Under the assumption that the three interferometers of the triangular configuration can become unavailable independently, or undergo staggered maintenance while the others remain in operation, the triangle can retain two operating V-shaped detectors when one is offline. It would then preserve sensitivity to both GW polarizations, whereas the loss of one detector in the 2L configuration would leave a single interferometer and substantially degrade the reconstruction of extrinsic parameters. Within this idealized availability model, two operating V-shaped detectors can outperform a single L-shaped interferometer for quantities such as luminosity distance and source-frame masses, even when the latter provides a larger SNR~\cite{Negri:2026gwp}.

However, whether this nominal redundancy translates into a larger effective duty cycle is an open operational question. If access to the triangular facility or maintenance of an individual beam line were to require suspending science operation in all three interferometers, their downtimes would be correlated and the assumed advantage would be reduced or absent. Requiring all six beam lines of the triangle to operate simultaneously may also introduce commissioning and reliability challenges distinct from those of the two geographically separated L-shaped detectors. A realistic comparison therefore requires facility-specific models of commissioning, maintenance access, failure correlations, recovery times, and partial-detector operation. The results of Ref.~\cite{Negri:2026gwp} should consequently be interpreted as illustrating the potential scientific value of the triangle's redundancy under its assumed availability model, rather than as demonstrating that the triangular configuration will necessarily achieve a higher duty cycle.\footnote{The degree of independence will also depend on the adopted infrastructure
concept. Designs in which suspension systems are hosted in mechanically and
acoustically isolated wells may, for example, facilitate maintenance of one
instrument while limiting disturbances to the others, whereas configurations
sharing larger underground volumes may exhibit stronger operational coupling.}

Our study addresses only one component of the broader ET site-selection and geometry problem: the propagation of representative site-dependent low-frequency sensitivity curves into a restricted set of science observables under idealized operating conditions. It should not be interpreted as an overall ranking of the candidate sites or as a definitive comparison between the triangular and 2L designs. A complete assessment must also include noise correlations and transient disturbances, detector duty cycle and maintenance, polarization reconstruction, stochastic-background sensitivity, calibration, null-stream performance, robustness against local environmental noise, construction and infrastructure constraints, costs, schedule, technological risk, and complementarity with other future observatories.

Different science cases weight these ingredients differently, and no single quantity --~such as SNR, arm length, low-frequency ASD, localization area, or observing uptime~-- is sufficient to characterize the overall scientific return. The final choice should therefore be informed by a multidimensional metric combining sensitivity, network geometry, operational reliability, noise independence, and the breadth of the accessible science program.

Within this broader framework, our results support one robust conclusion: preserving the best possible sensitivity below approximately $10\,{\rm Hz}$ is essential for realizing the distinctive scientific potential of ET. The advantage of longer arms or of a particular detector geometry cannot be assessed independently of the noise that can realistically be achieved at the selected sites. Continued site characterization, realistic end-to-end noise simulations, and science studies including correlated and non-Gaussian noise, time-dependent detector availability, broader source populations, and more complete waveform models (e.g., including eccentricity and spin precession, whose signatures are especially sensitive to the detector performance at low frequencies) will therefore be crucial for translating the ET design choices into a reliable assessment of its ultimate scientific performance.

\section*{Acknowledgments}
We thank Ulyana Dupletsa and Filippo Santoliquido for useful discussions, as well as Riccardo DeSalvo, Krishnendu Naderi Varium, Mariusz Suchenek, and Valentin Tempel for comments on the draft. This work is supported by the MUR FIS2 Advanced Grant ET-NOW (CUP:~B53C25001080001) and the INFN TEONGRAV initiative.
The authors gratefully acknowledge the financial support provided by the German Federal Ministry of Research, Technology and Space (BMFTR) in the framework of the ``Knowledge creates perspectives for the region!'', for the project StStG – DZA – Aufbauphase: Deutsches Zentrum für Astrophysik, Großforschungszentrum in der sächsischen Lausitz: Aufbauphase 2026, grant number 03WSP1746.
Numerical computations have been performed at the Vera and CHRONOS clusters supported by the Italian Ministry of Research and by Sapienza University of Rome.

\section{Appendix}
In this appendix, we present supplementary results from the Bayesian inference analyses discussed in the main text.

Table~\ref{tab:IMBH_priors} summarizes the priors adopted for the intermediate-mass BH analyses. We sample directly over the detector-frame chirp mass and symmetric mass ratio, while imposing constraint priors on the derived detector-frame component masses. The luminosity-distance prior is uniform in comoving volume, with bounds adapted to the injected value, whereas the remaining extrinsic parameters are assigned the broad priors specified in the table.

\begin{figure}[t]
    \centering
    \includegraphics[width=0.48\linewidth]{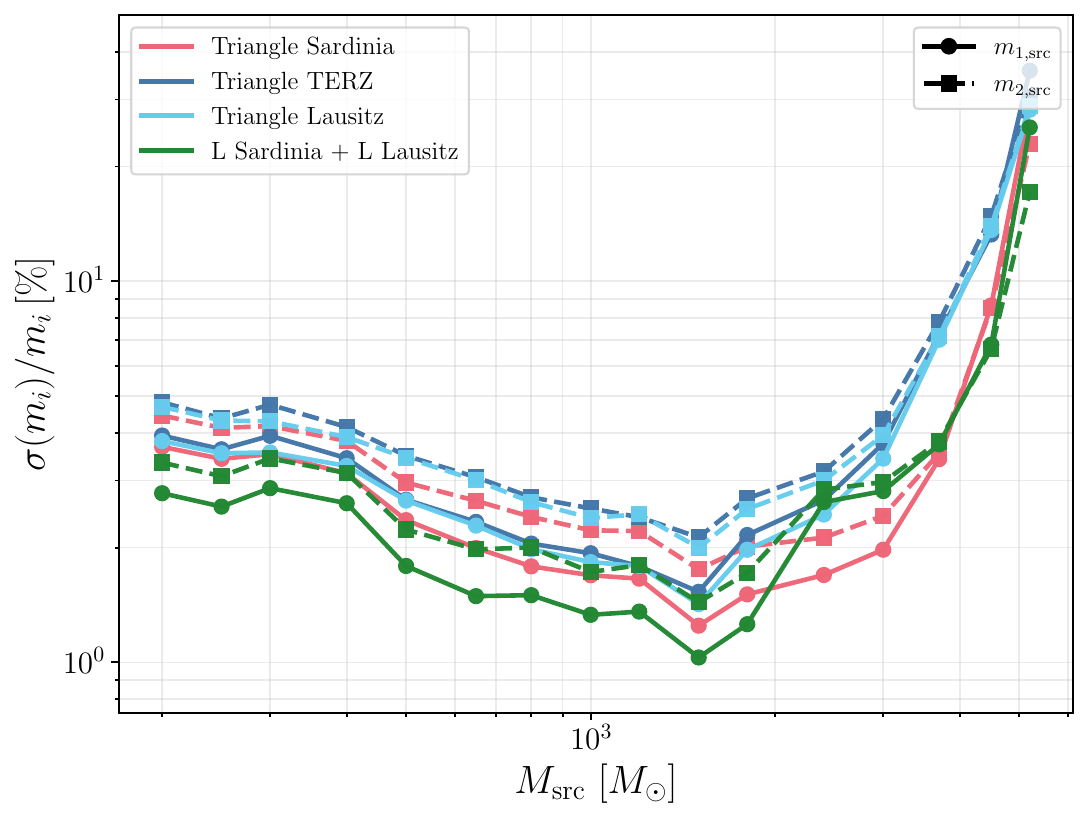}
    \includegraphics[width=0.48\linewidth]{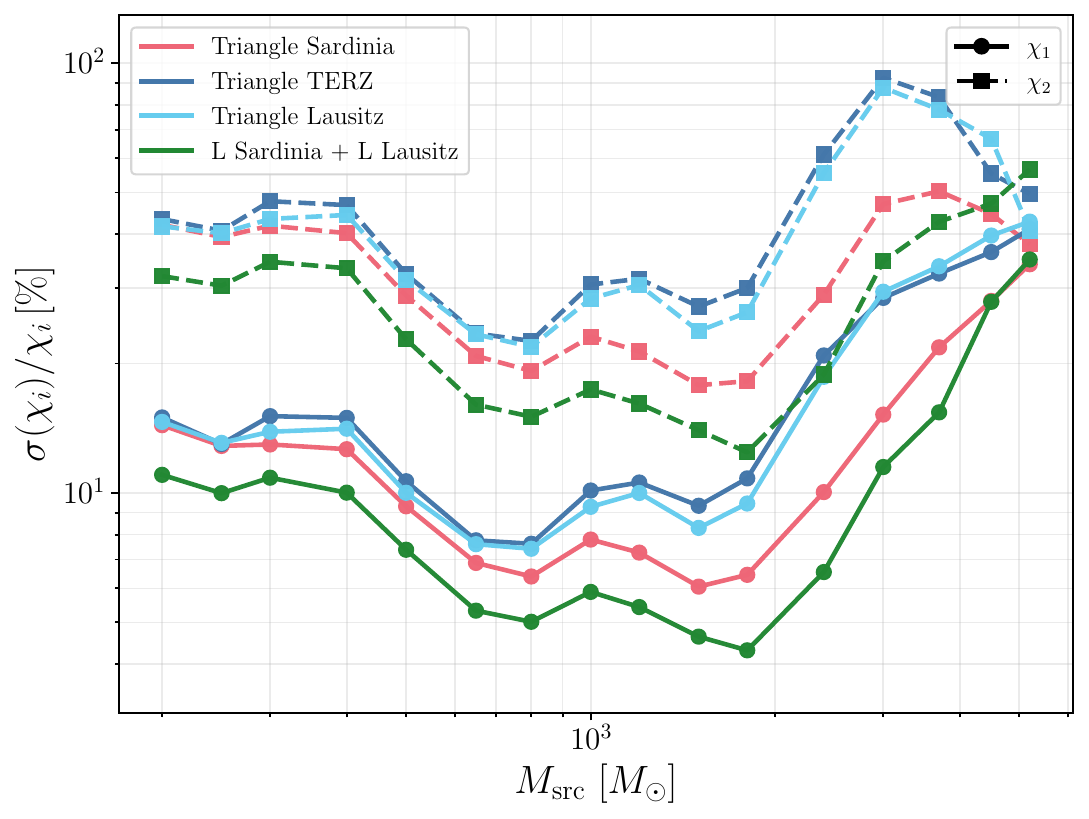}
    
    \caption{Left panel: relative uncertainties on the source-frame component masses as functions of the injected source-frame total mass, for the four detector configurations considered. Right panel: corresponding relative uncertainties on the individual spins. In both panels, solid and dashed curves refer to the primary and secondary components, respectively.}
    \label{fig:masses_spins_source}
    
\end{figure}

Figure~\ref{fig:masses_spins_source} presents additional results for the source-frame component masses and individual spins. At $z=1$, the qualitative hierarchy among detector configurations is consistent with that discussed in the main text:  the TERZ and Lausitz triangles yield broadly comparable uncertainties, with a modest tendency for Lausitz to perform better, the Sardinia triangle performs overall better among the three triangle configurations, and the 2L configuration generally provides the tightest constraints. The differences become more pronounced at larger source masses, as an increasing fraction of the signal shifts into the low-frequency band.

There is, however, a range around $M_{\rm src}\sim {\rm few}\times10^3\,M_{\odot}$ in which the source-frame mass uncertainties obtained with the 2L configuration are larger than those of the triangular configurations, despite the higher SNR of the 2L network. This degradation originates from multimodal structures in the luminosity-distance posterior, caused by residual degeneracies between distance and sky position in the 2L configuration. The resulting uncertainty in $d_L$, and hence in the inferred redshift, propagates directly to the source-frame masses through $M_{\rm src}=M_{\rm det}/(1+z)$. Such multimodalities are expected for sufficiently massive BH binaries, whose signals remain in the detector band for only a short time.

\begin{figure*}[t]
    \centering
    \includegraphics[width=\textwidth]{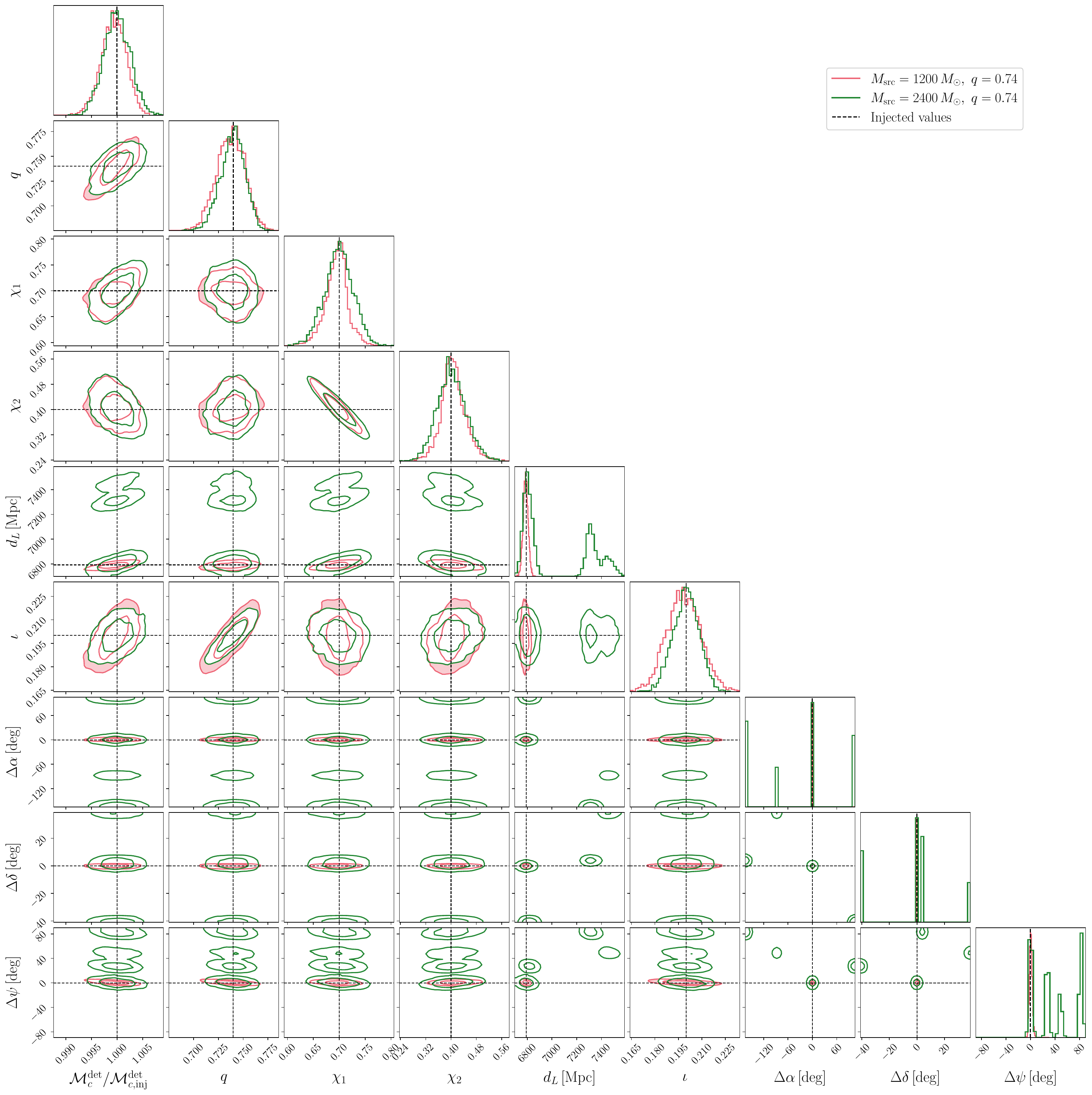}
    \caption{
        Comparison of the posterior distributions for two IMBH systems with source-frame total masses $M_{\rm src}=1200\,M_\odot$ and $M_{\rm src}=2400\,M_\odot$, both with mass ratio $q=0.74$, observed with the 2L configuration. The detector-frame chirp mass is normalized to its injected value, while the right ascension, declination, and polarization angle are shown as offsets from their respective injected values. Dashed black lines mark the injected parameters. The one-dimensional marginalized posteriors are normalized to unit peak for ease of comparison, and the contours enclose the $50\%$ and $90\%$ credible regions.
    }
    \label{fig:imbh_corner_mass_comparison}
\end{figure*}

This effect has been investigated in detail in Refs.~\cite{Santoliquido:2025lot,Santoliquido:2025aiq}. Figure~\ref{fig:imbh_corner_mass_comparison} illustrates it for two injections with different source-frame total masses: $M_{\rm src}=1200\,M_{\odot}$, for which the luminosity-distance posterior is essentially unimodal, and $M_{\rm src}=2400\,M_{\odot}$, for which a secondary mode appears. In the latter case, the secondary peak is located at approximately $7400\,{\rm Mpc}$, more than $3\sigma$ away from the injected value $d_{\rm L,inj}=6792\,{\rm Mpc}$. The same degeneracy also produces multiple modes in the sky-position and polarization-angle posteriors. By contrast, the spin measurements are not directly affected, since they do not depend on the redshift conversion. Overall, as shown in the right panel of Fig.~\ref{fig:masses_spins_source}, the individual spins are measured significantly less accurately than the masses, especially for the secondary component.

\begin{table}[t]
\centering
\begin{tabular}{lll}
\toprule
Parameter & Prior & Notes \\
\midrule
\multicolumn{3}{c}{Intrinsic parameters} \\
\midrule

\(\mathcal{M}^{\rm det}\) &
\(\mathcal{U}\!\left[
0.6\,\mathcal{M}^{\rm det}_{\rm inj},
1.4\,\mathcal{M}^{\rm det}_{\rm inj}
\right]\) &
Detector-frame chirp mass \\

\(\eta\) &
\(\mathcal{U}(0,0.25)\) &
Symmetric mass ratio \\

\(m_1^{\rm det}\) &
\(\mathcal{C}(5\,M_\odot,m_{\max})\) &
Constraint on the derived primary mass \\

\(m_2^{\rm det}\) &
\(\mathcal{C}(5\,M_\odot,m_{\max})\) &
Constraint on the derived secondary mass \\

\(\chi_1\) &
\(\mathcal{U}(0,0.97)\) &
Primary aligned-spin parameter \\

\(\chi_2\) &
\(\mathcal{U}(0,0.97)\) &
Secondary aligned-spin parameter \\

\midrule
\multicolumn{3}{c}{Extrinsic parameters} \\
\midrule

\(d_L\) &
\(\mathcal{U}_{V_c}(d_{L,\min},d_{L,\max})\) &
Uniform in comoving volume \\

\(\iota\) &
$\propto\sin\iota$,
\(\iota\in[0,\pi]\) &
Isotropic binary orientation \\

\(t_c\) &
\(\mathcal{U}(t_{c,\rm inj}-0.1\,{\rm s},
t_{c,\rm inj}+0.1\,{\rm s})\) &
Geocentric coalescence time \\

\(\alpha\) &
\(\mathcal{U}(0,2\pi)\) &
Right ascension \\

\(\delta\) &
$\propto\cos\delta$,
\(\delta\in[-\pi/2,\pi/2]\) &
Isotropic sky position \\

\(\psi\) &
\(\mathcal{U}(0,\pi)\) &
Polarization angle \\

\(\phi_c\) &
\(\mathcal{U}(0,2\pi)\) &
Coalescence phase \\

\bottomrule
\end{tabular}

\caption{
Priors adopted in the Bayesian inference analysis of intermediate-mass
black-hole binaries. Here, \(\mathcal{U}\) denotes a uniform prior,
\(\mathcal{U}_{V_c}\) a prior uniform in comoving volume, and
\(\mathcal{C}\) a constraint prior. The detector-frame component masses
are not sampled directly: the sampling is performed over
\(\mathcal{M}^{\rm det}\) and \(\eta\), while constraints are imposed on
the corresponding derived values of \(m_1^{\rm det}\) and
\(m_2^{\rm det}\). The luminosity-distance bounds are $d_{L,\min}
=
\max\!\left[
10^2\,{\rm Mpc},
\kappa_{d_L}^{\rm low}d_{L,\rm inj}
\right],
\qquad
d_{L,\max} = \kappa_{d_L}^{\rm high}d_{L,\rm inj}$, with
$\kappa_{d_L}^{\rm low}=0.5$ and
$\kappa_{d_L}^{\rm high}=2.5$.
The upper component-mass bound is $m_{\max}
= \max\!\left[
10^3\,M_\odot,
\kappa_m
\max\!\left(
m_{1,\rm inj}^{\rm det},
m_{2,\rm inj}^{\rm det}
\right)
\right]$ with $\kappa_m=2.5$.
}
\label{tab:IMBH_priors}
\end{table}

Table~\ref{tab:BNS_priors} summarizes the priors adopted in the analysis for the BNS runs. As in the case of the IMBHs, we sample directly over the detector-frame chirp mass and symmetric mass ratio, while imposing constraint priors on the detector-frame component masses, with the only difference that the ranges are reduced to speed up the computation. The luminosity-distance prior is uniform in comoving volume, with bounds adapted to the injected distance, whereas the remaining extrinsic parameters are assigned the standard broad priors described in the table. 

\begin{table*}[t]
\centering
\begin{tabular}{lll}
\toprule
Parameter & Prior & Notes \\
\midrule
\multicolumn{3}{c}{Intrinsic parameters} \\
\midrule

\(\mathcal{M}^{\rm det}\) &
\(\mathcal{U}\!\left[
0.999\,\mathcal{M}^{\rm det}_{\rm inj},
1.001\,\mathcal{M}^{\rm det}_{\rm inj}
\right]\) &
Detector-frame chirp mass \\

\(\eta\) &
\(\mathcal{U}(0.1,0.25)\) &
Symmetric mass ratio \\

\(m_1^{\rm det}\) &
\(\mathcal{C}(0.5\,M_\odot,m_{\max})\) &
Constraint on the derived primary mass \\

\(m_2^{\rm det}\) &
\(\mathcal{C}(0.5\,M_\odot,m_{\max})\) &
Constraint on the derived secondary mass \\

\(\chi_1\) &
\(\pi_{\rm AS}(\chi_1;a_{\max}=0.1)\) &
Primary aligned-spin component \\

\(\chi_2\) &
\(\pi_{\rm AS}(\chi_2;a_{\max}=0.1)\) &
Secondary aligned-spin component \\

\(\Lambda_1\) &
\(\delta(\Lambda_1-\Lambda_{1,\rm inj})\) &
Fixed primary tidal deformability \\

\(\Lambda_2\) &
\(\delta(\Lambda_2-\Lambda_{2,\rm inj})\) &
Fixed secondary tidal deformability \\

\midrule
\multicolumn{3}{c}{Extrinsic parameters} \\
\midrule

\(d_L\) &
\(\mathcal{U}_{V_c}(d_{L,\min},d_{L,\max})\) &
Uniform in comoving volume \\

\(\iota\) &
\(\propto\sin\iota\),
\(\iota\in[0,\pi]\) &
Isotropic binary orientation \\

\(t_c\) &
\(\mathcal{U}\!\left(
t_{c,\rm inj}-0.1\,{\rm s},
t_{c,\rm inj}+0.1\,{\rm s}
\right)\) &
Geocentric coalescence time \\

\(\alpha\) &
\(\mathcal{U}(0,2\pi)\) &
Right ascension \\

\(\delta\) &
\(\propto\cos\delta\),
\(\delta\in[-\pi/2,\pi/2]\) &
Isotropic sky position \\

\(\psi\) &
\(\mathcal{U}(0,\pi)\) &
Polarization angle \\

\(\phi_c\) &
\(\mathcal{U}(0,2\pi)\) &
Coalescence phase \\

\bottomrule
\end{tabular}

\caption{
Priors adopted in the Bayesian inference analysis of binary neutron-star
signals. Here, \(\mathcal{U}\) denotes a uniform prior,
\(\mathcal{U}_{V_c}\) a prior uniform in comoving volume,
\(\mathcal{C}\) a constraint prior, and \(\delta\) a Dirac delta.
The detector-frame component masses are not sampled directly: sampling is
performed over the detector-frame chirp mass
\(\mathcal{M}^{\rm det}\) and the symmetric mass ratio \(\eta\), while
constraints are imposed on the corresponding derived component masses.
The luminosity-distance bounds are
$d_{L,\min}
=
\max\!\left[
10^{-3}\,{\rm Mpc},
\kappa_{d_L}^{\rm low}d_{L,\rm inj}
\right],
\qquad
d_{L,\max}
=
\kappa_{d_L}^{\rm high}d_{L,\rm inj},$
with $\kappa_{d_L}^{\rm low}=0.5$ and
$\kappa_{d_L}^{\rm high}=2.5$.
The upper component-mass bound is
$m_{\max}
=
\max\!\left[
3\,M_\odot,
\kappa_m
\max\!\left(
m_{1,\rm inj}^{\rm det},
m_{2,\rm inj}^{\rm det}
\right)
\right],$
with $\kappa_m=2.5$.
For each aligned-spin component,
$\chi_i=a_i\cos\theta_i$, with
$a_i\sim\mathcal{U}(0,0.1)$ and
$\cos\theta_i\sim\mathcal{U}(-1,1)$; the resulting prior is denoted by
$\pi_{\rm AS}$.
The tidal deformabilities are fixed to their injected values during parameter
estimation.
}
\label{tab:BNS_priors}
\end{table*}

Table~\ref{tab:solid_angles_theta} reports the same results of Table~\ref{tab:solid_angles}, but for the prior of $\iota$ in the range $[\pi/2,\pi]$, thus forcing the absence of multimodalities.

\begin{table}[htbp]
    \centering
    \begin{tabular}{lccc}
        \hline
        Configuration
        & \multicolumn{3}{c}{$\Delta\Omega_{90\%}\,[\mathrm{deg}^2]$} \\
        \cline{2-4}
        & $3\,{\rm h}$
        & $1\,{\rm h}$
        & $30\,{\rm min}$ \\
        \hline
        Triangle TERZ            & $62.6$ & $2.00$ & $0.52$ \\
        Triangle Lausitz         & $44.6$ & $1.49$ & $0.48$ \\
        Triangle Sardinia        & $5.96$ & $0.64$ & $0.32$ \\
        L Sardinia + L Lausitz   & $3.46$ & $0.32$ & $0.16$ \\
        \hline
    \end{tabular}
    \caption{Same as Table~\ref{tab:solid_angles}, but for the prior of $\iota\in[\pi/2,\pi]$.}
    \label{tab:solid_angles_theta}
\end{table}

\begin{figure*}[htbp]
    \centering

    \begin{minipage}[c]{0.44\textwidth}
        \centering
        \includegraphics[width=\textwidth]{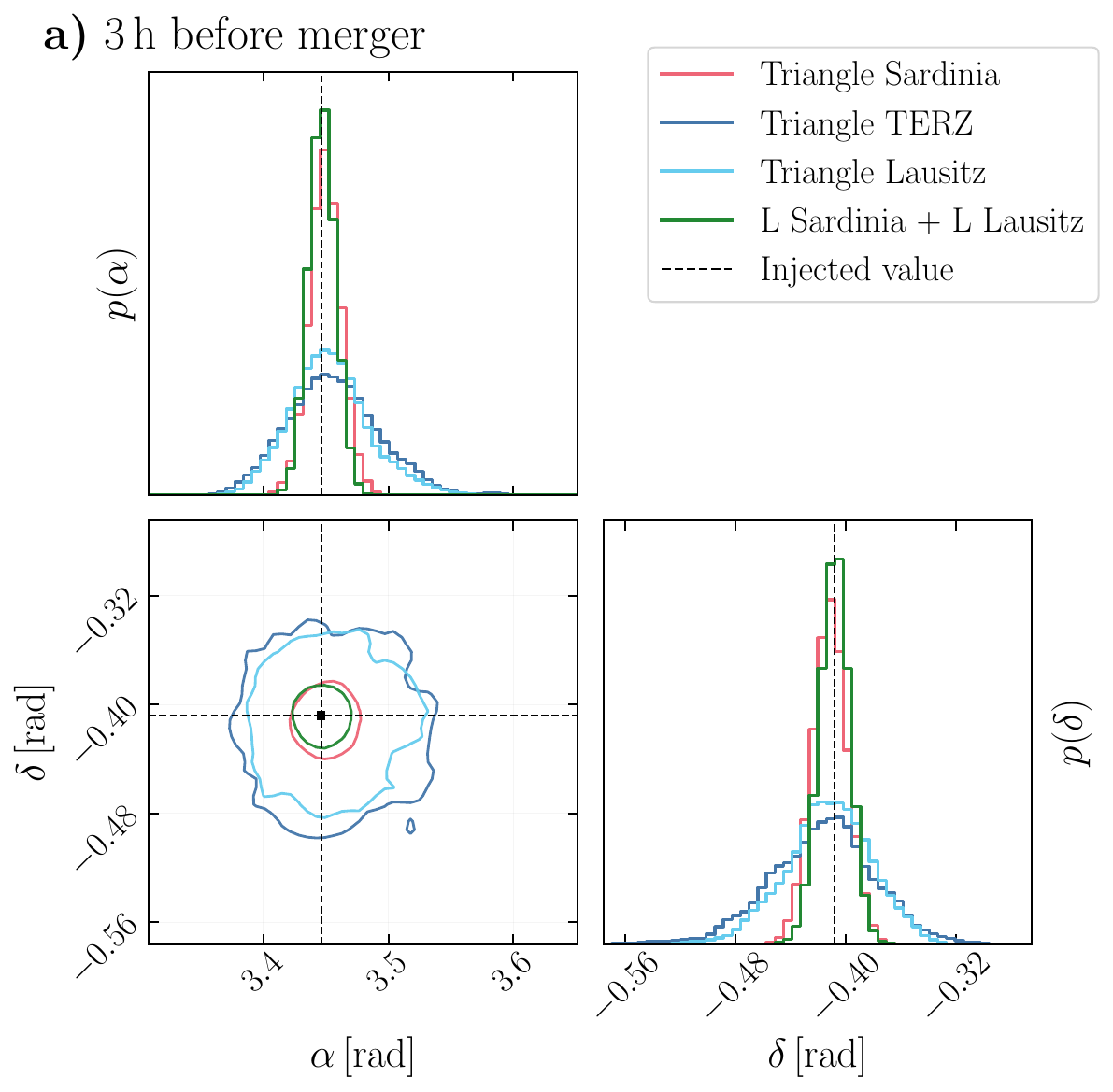}
    \end{minipage}
    \hfill
    \begin{minipage}[c]{0.42\textwidth}
        \centering
        \includegraphics[width=\textwidth]{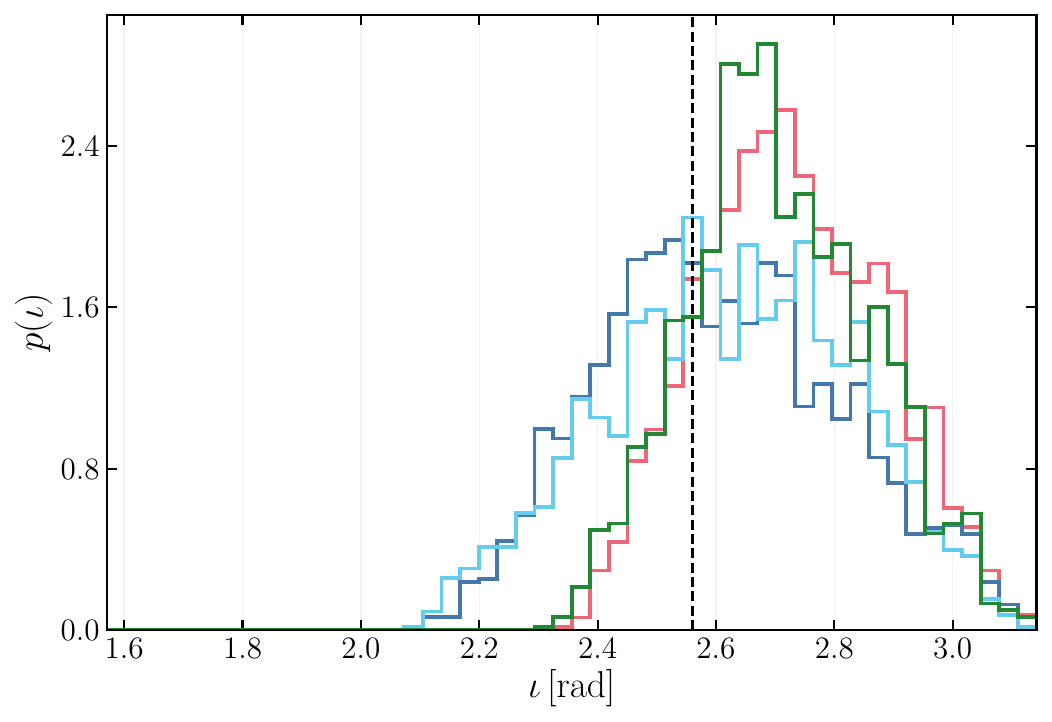}
    \end{minipage}

    \vspace{0.25cm}

    \begin{minipage}[c]{0.44\textwidth}
        \centering
        \includegraphics[width=\textwidth]{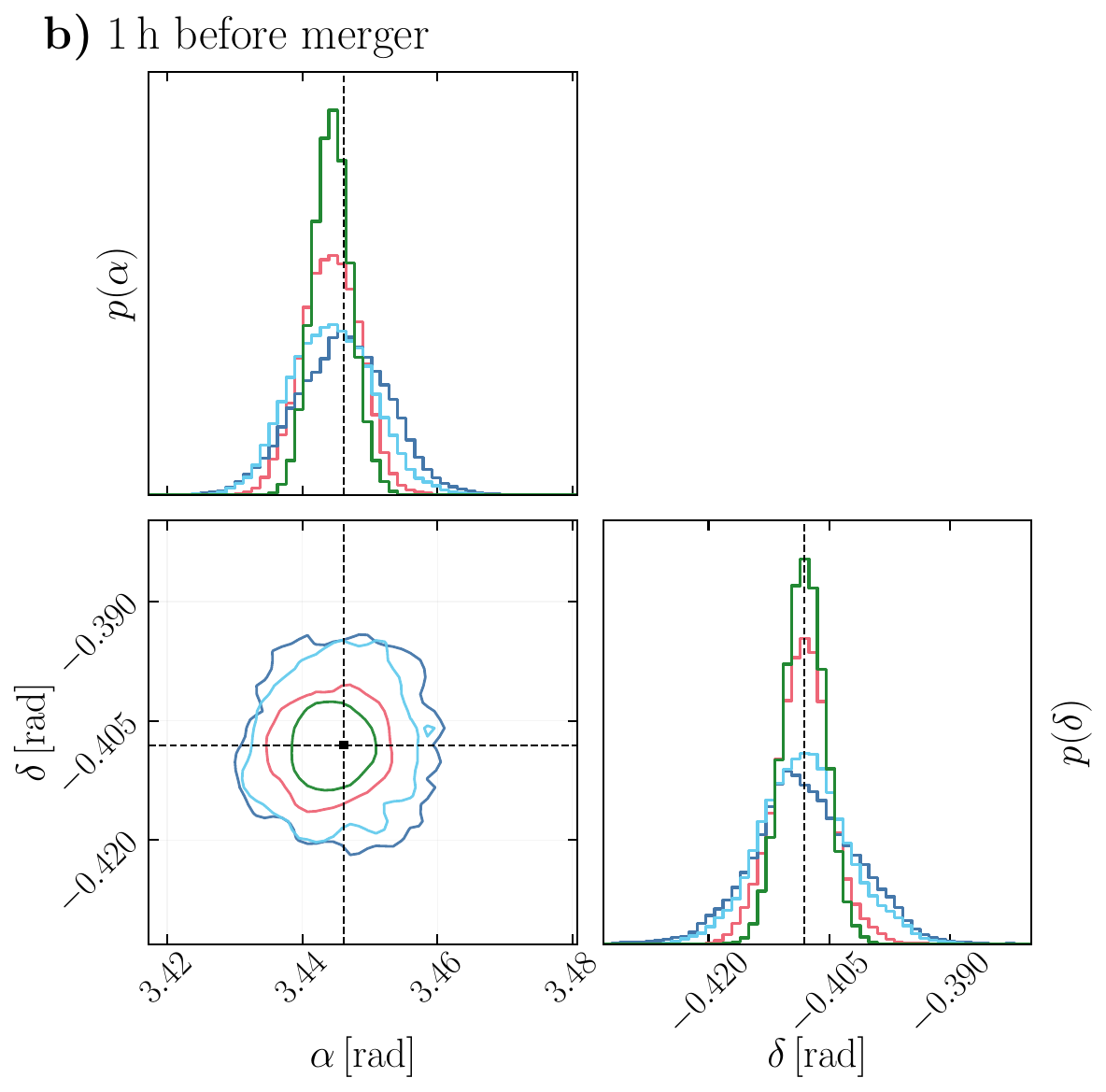}
    \end{minipage}
    \hfill
    \begin{minipage}[c]{0.42\textwidth}
        \centering
        \includegraphics[width=\textwidth]{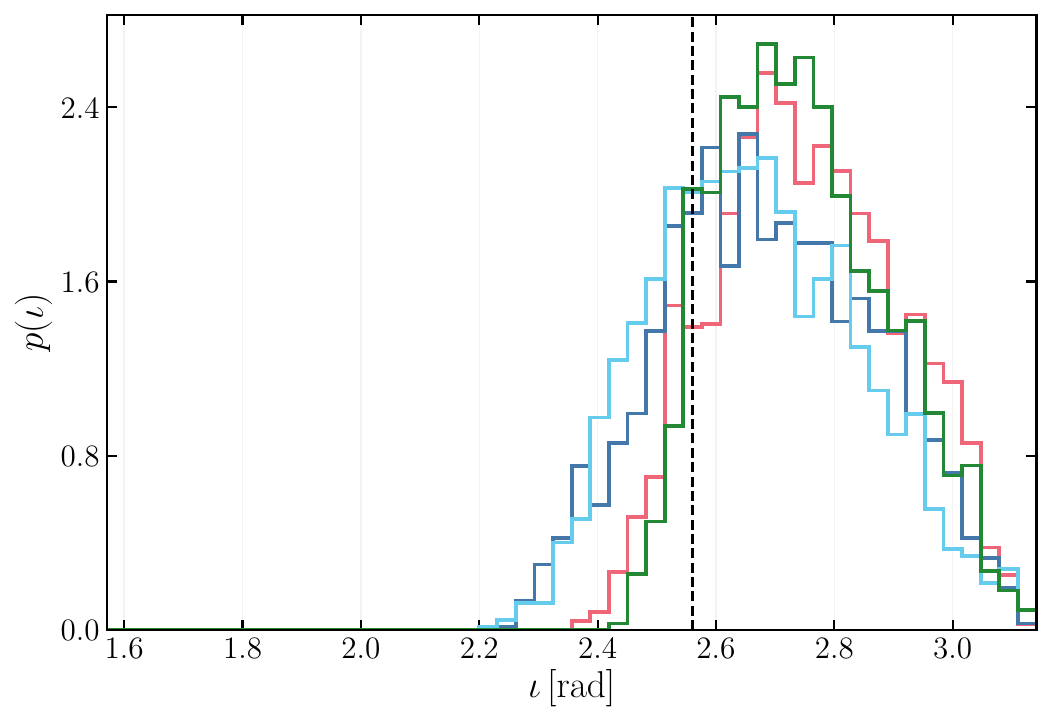}
    \end{minipage}

    \vspace{0.25cm}

    \begin{minipage}[c]{0.44\textwidth}
        \centering
        \includegraphics[width=\textwidth]{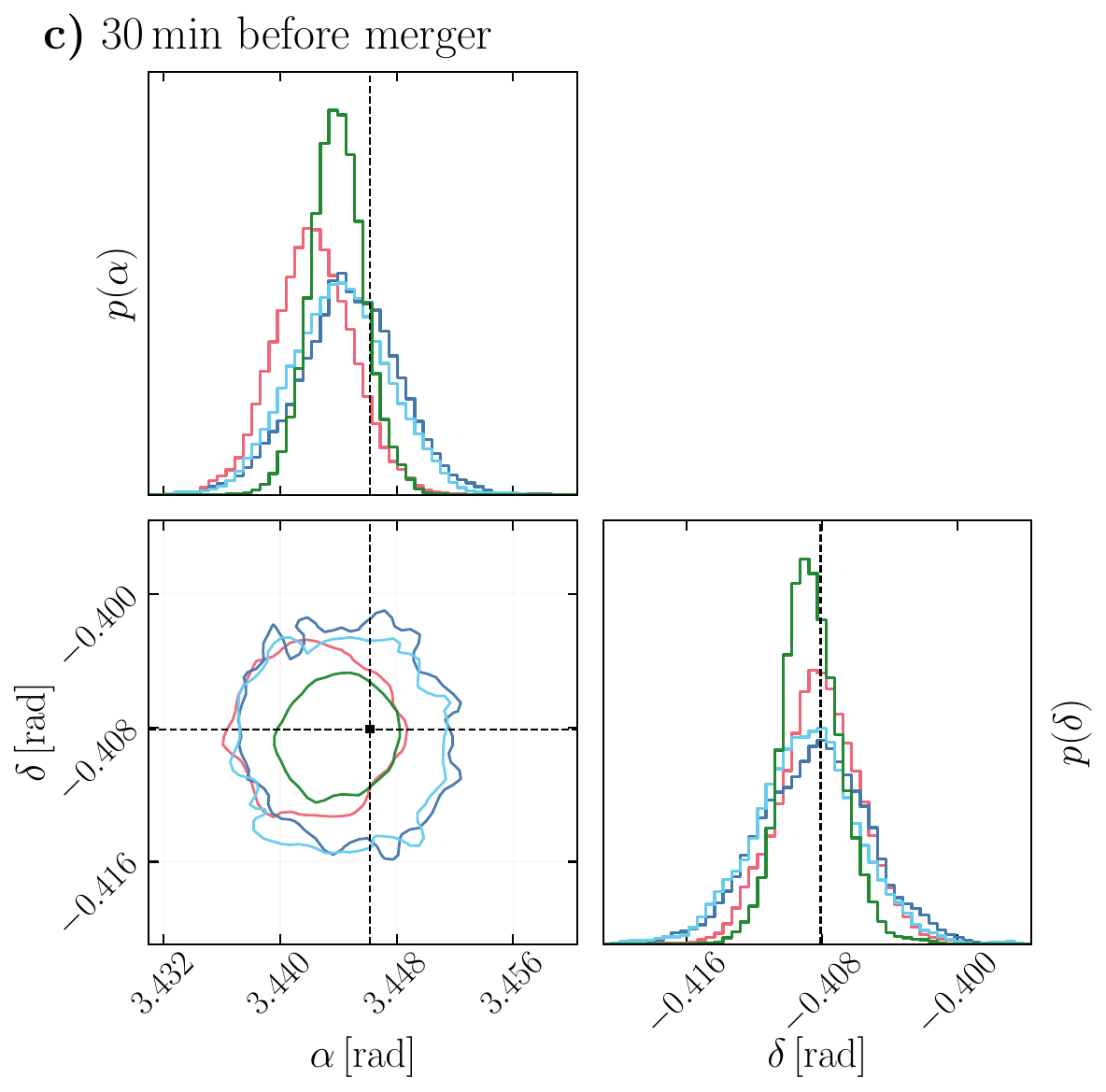}
    \end{minipage}
    \hfill
    \begin{minipage}[c]{0.42\textwidth}
        \centering
        \includegraphics[width=\textwidth]{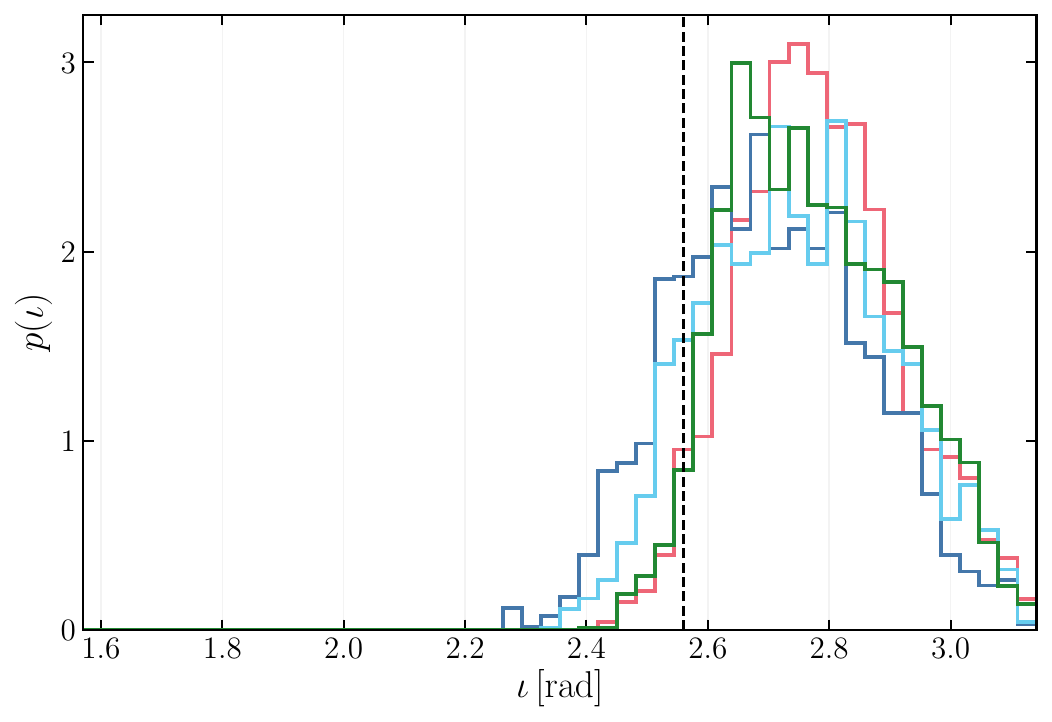}
    \end{minipage}

    \caption{Same as Fig.~\ref{fig:posteriors_BNS_standard} but imposing $\pi/2 \leq \iota \leq \pi$ as a prior, thus forcing the absence of multimodalities. Panels a), b), and c) correspond to $T=10800\,\mathrm{s}$, $T=3600\,\mathrm{s}$, and $T=1800\,\mathrm{s}$ before merger, respectively. In each panel, the left plot shows the joint and marginalized posterior distributions of right ascension $\alpha$ and declination $\delta$, while the right plot shows the one-dimensional posterior probability density $p(\iota)$ of the inclination angle.}
    \label{fig:posteriors_BNS_theta_geq_pi2}
\end{figure*}

Finally, Fig.~\ref{fig:posteriors_BNS_theta_geq_pi2} shows the joint posterior distributions of the sky position and inclination angle at $3\mathrm{h}$, $1\mathrm{h}$, and $30\mathrm{min}$ before merger, respectively, when instead sampling over $\iota\in[\pi/2,\pi]$.

\clearpage

\bibliographystyle{JHEP}
\bibliography{main}

\end{document}